\documentstyle[12pt]{article}

\def\hybrid{\topmargin -20pt  \oddsidemargin 0pt
      \headheight 0pt   \headsep 0pt
      \textwidth 6.25in % A4 paper
      \textheight 9.5in % A4 paper
      \marginparwidth .875in
      \parskip 5pt plus 1pt   \jot = 1.5ex}

\hybrid

\begin{document}
%\titlepage
\def\x{\times}
\def\beq{\begin{equation}}
\def\eeq{\end{equation}}
\def\beqa{\begin{eqnarray}}
\def\eeqa{\end{eqnarray}}
\def\L{ {\cal L}}
\def\C{ {\cal C}}
\def\N{ {\cal N}}
\def\calE{{\cal E}}
\def\lin{{\rm lin}}
\def\Tr{{\rm Tr}}
\def\cF{{\cal F}}
\def\cD{{\cal D}}
\def\modS{{S+\bar S}}
\def\mods{{s+\bar s}}
\newcommand{\Fg}[1]{{F}^{({#1})}}
\newcommand{\cFg}[1]{{\cal F}^{({#1})}}
\newcommand{\cFgc}[1]{{\cal F}^{({#1})\,{\rm cov}}}
\newcommand{\Fgc}[1]{{F}^{({#1})\,{\rm cov}}}
\def\mpl{m_{\rm Planck}}
\def\mxth{\mathsurround=0pt }
\def\xversim#1#2{\lower2.pt\vbox{\baselineskip0pt \lineskip-.5pt
x  \ialign{$\mxth#1\hfil##\hfil$\crcr#2\crcr\sim\crcr}}}
\def\simgr{\mathrel{\mathpalette\xversim >}}
\def\simle{\mathrel{\mathpalette\xversim <}}

\newcommand{\ms}[1]{\mbox{\scriptsize #1}}
\renewcommand{\a}{\alpha}
\renewcommand{\b}{\beta}
\renewcommand{\c}{\gamma}
\renewcommand{\d}{\delta}
\newcommand{\th}{\theta}
\newcommand{\TH}{\Theta}
\newcommand{\pa}{\partial}
\newcommand{\g}{\gamma}
\newcommand{\G}{\Gamma}
\newcommand{\A}{\Alpha}
\newcommand{\B}{\Beta}
\newcommand{\D}{\Delta}
\newcommand{\e}{\epsilon}
\newcommand{\E}{{\cal E}}
\newcommand{\z}{\zeta}
\newcommand{\Z}{\Zeta}
\newcommand{\k}{\kappa}
\newcommand{\K}{\Kappa}
\renewcommand{\l}{\lambda}
\renewcommand{\L}{\Lambda}
\newcommand{\m}{\mu}
\newcommand{\M}{\Mu}
\newcommand{\n}{\nu}
\newcommand{\X}{\Chi}
\newcommand{\R}{\Rho}
\newcommand{\s}{\sigma}
\renewcommand{\S}{\Sigma}
\renewcommand{\t}{\tau}
\newcommand{\T}{\Tau}
\newcommand{\y}{\upsilon}
\newcommand{\Y}{\upsilon}
\renewcommand{\o}{\omega}
\newcommand{\q}{\theta}
\newcommand{\h}{\eta}

\def\dota{ {\dot{\alpha}} }
\def\lag{Lagrangian}
\def\Kahler{K\"{a}hler}
\def\kahler{K\"{a}hler}
\def\A{ {\cal A}}
\def\F{{\cal F}}
\def\cL{ {\cal L}}

\def\R{ {\cal R}}
\def\x{ \times }
\def\beq{\begin{equation}}
\def\eeq{\end{equation}}
\def\beqa{\begin{eqnarray}}
\def\eeqa{\end{eqnarray}}

\sloppy
\newcommand{\be}{\begin{equation}}
\newcommand{\eq}{\end{equation}}
\newcommand{\ov}{\overline}
\newcommand{\un}{\underline}
\newcommand{\p}{\partial}
\newcommand{\la}{\langle}
\newcommand{\ra}{\rangle}
\newcommand{\bl}{\boldmath}
\newcommand{\ds}{\displaystyle}
\newcommand{\nl}{\newline}
\newcommand{\Nzahl}{{\bf N}  }
\newcommand{\zzahl}{ {\bf Z} }
\newcommand{\Zzahl}{ {\bf Z} }
\newcommand{\Qzahl}{ {\bf Q}  }
\newcommand{\Rzahl}{ {\bf R} }
\newcommand{\Czahl}{ {\bf C} }
\newcommand{\wt}{\widetilde}
\newcommand{\wh}{\widehat}
\newcommand{\fs}[1]{\mbox{\scriptsize \bf #1}}
\newcommand{\ft}[1]{\mbox{\tiny \bf #1}}
\newtheorem{satz}{Satz}[section]
\newenvironment{Satz}{\begin{satz} \sf}{\end{satz}}
\newtheorem{definition}{Definition}[section]
\newenvironment{Definition}{\begin{definition} \rm}{\end{definition}}
\newtheorem{bem}{Bemerkung}
\newenvironment{Bem}{\begin{bem} \rm}{\end{bem}}
\newtheorem{bsp}{Beispiel}
\newenvironment{Bsp}{\begin{bsp} \rm}{\end{bsp}}
\renewcommand{\arraystretch}{1.5}

%\textwidth14.5cm
%\textheight23.0cm
%\oddsidemargin0.5cm
%\topmargin-1.4cm

%\addtocounter{section}{1}

\renewcommand{\thesection}{\arabic{section}}
\renewcommand{\theequation}{\thesection.\arabic{equation}}

%\setcounter{section}{1}
%\addtocounter{section}{1}
\parindent0em

\def\S4{\frac{SO(4,2)}{SO(4) \otimes SO(2)}}
\def\P3{\frac{SO(3,2)}{SO(3) \otimes SO(2)}}
\def\MGd{\frac{SO(r,p)}{SO(r) \otimes SO(p)}}
\def\SOd{\frac{SO(r,2)}{SO(r) \otimes SO(2)}}
\def\SO2{\frac{SO(2,2)}{SO(2) \otimes SO(2)}}
\def\SUm{\frac{SU(n,m)}{SU(n) \otimes SU(m) \otimes U(1)}}
\def\SUS{\frac{SU(n,1)}{SU(n) \otimes U(1)}}
\def\SK{\frac{SU(2,1)}{SU(2) \otimes U(1)}}
\def\SU{\frac{ SU(1,1)}{U(1)}}

\begin{titlepage}
\begin{center}
\hfill CERN-TH/96-227\\
\hfill HUB-EP-96/47\\
\hfill {\tt hep-th/9608154}\\

\vskip .6in

{\bf PERTURBATIVE COUPLINGS AND MODULAR FORMS
 IN $N=2$ STRING MODELS WITH A WILSON LINE}

\vskip .2in

{\bf 
Gabriel Lopes Cardoso$^a$, Gottfried Curio$^b$,
Dieter L\"ust$^b$}\footnote{email: \tt 
cardoso@surya11.cern.ch,
curio@qft2.physik.hu-berlin.de,\\			
		luest@qft1.physik.hu-berlin.de}
\\
\vskip 1.2cm

$^a${\em Theory Division, CERN, CH-1211 Geneva 23, Switzerland}\\
$^b${\em Humboldt-Universit\"at zu Berlin,
Institut f\"ur Physik, 
D-10115 Berlin, Germany}

\vskip .1in

\end{center}

\vskip .2in

\begin{center} {\bf ABSTRACT } \end{center}
\begin{quotation}\noindent

We consider a class of four parameter $D=4, N=2$ 
string models, namely
heterotic strings compactified on
$K_3\times T_2$
together with their dual type II partners
on  Calabi-Yau three-folds.  
With the help of generalized modular forms (such
as Siegel and Jacobi forms), we compute the perturbative
prepotential and the perturbative Wilsonian gravitational
coupling $F_1$ for each of the models in this class.
We check heterotic/type II duality for one of the models by
relating the modular forms in the heterotic description
to the known instanton numbers in the type II description.
We comment on the relation of our results
to recent proposals for closely related models.

\end{quotation}
\vskip 2.5cm
August 1996\\
\hfill CERN-TH/96-227\\
\end{titlepage}
\vfill
\eject

%%%%%%%%%%%%%%%%%%%%%%%%%%%%%%%%%%%%%%%%%%%%%%%%%%%%%%%%%%%%%
\newpage

\section{Introduction}

Recently, accumulating evidence for the existence of various types
of strong--weak coupling duality symmetries was gathered, such as 
$S$-duality of the four-dimensional $N=4$ heterotic string 
\cite{sduality,SchSen,Sen} and
string-string dualities between the heterotic and type II strings
\cite{HullTown,Wit1,KV}.
The 
string-string duality between
four-dimensional strings with $N=2$ space-time supersymmetry \cite{KV} is
of particular interest, since $N=2$ strings exhibit a very rich
non-perturbative structure which, in the point particle limit,
contains \cite{KKLMV} 
the non-perturbative effects of rigid $N=2$ gauge theories
\cite{SW}. Furthermore, the $N=2$ strings are ``half way" in between the
well controlled $N=4$ models and the phenomenologically interesting, but
much less understood $N=1$ string-string dualities \cite{nonedual}.

The $N=2$ string-string duality between heterotic strings 
on $K3\times T_2$ and corresponding type II strings
on a suitably chosen Calabi--Yau three-fold has been successfully
tested \cite{KV},\cite{KLT}--\cite{CCLM}
for models with  a small number of vector multiplets. 
Most of these tests were based on the comparison of lower order gauge
and gravitational couplings \cite{WKLL,AFGNT,HM}
of the perturbative heterotic string with the corresponding couplings
of the dual type II string in some corner of the
Calabi--Yau moduli space.
One key point in establishing the string-string duality between heterotic
and type II $N=2$ strings is the appearance 
\cite{FLST} of certain modular functions
in the low-energy effective action of these theories.

To be more specific, the discussion so far was essentially limited
to models with number of massless Abelian vector multiplets $N_V=3$
and $N_V=4$. For the rank four case, $N_V=4$, 
one is dealing with the heterotic $S$-field,
with two $T_2$ moduli $T$ and $U$ plus the graviphoton. 
The perturbative heterotic vector multiplet couplings are given in terms
of modular functions of the perturbative $T$-duality group $SO(2,2;{\bf Z})$.
Due to the required
embedding of this $T$-duality group
into the $N=2$ symplectic transformations it follows \cite{WKLL,AFGNT} that
the heterotic one-loop prepotential must obey well-defined 
transformation rules under this
group. In addition it was shown in \cite{HM} that the one-loop prepotential 
can be expressed in terms of the coefficients of the $q$ expansion of certain
modular forms.
This heterotic $S$-$T$-$U$ model is supposed \cite{KV}
to be dual to the type II string
compactified on  the Calabi--Yau space $P_{1,1,2,8,12}(24)$ with $h_{1,1}=3$,
$h_{2,1}=243$.
In fact, it was shown for this example that the perturbative heterotic
prepotential and the function $F_1$ (which specifies the non-minimal
gravitational interactions involving the square of the Riemann tensor)
agree with the corresponding type II functions in the limit
where one specific K\"ahler class modulus of the underlying Calabi--Yau
space becomes large. A set of interesting relations between certain
topological Calabi--Yau data (rational and elliptic instanton numbers)
and various modular forms has emerged when performing these tests 
\cite{HM,CCLM}.

It is clearly an interesting problem to extend this kind
of discussion to $N=2$ string models with a larger number of vector
multiplets, $N_V>4$. It is the purpose of this paper to 
compute  the heterotic one-loop couplings
as well as to discuss the heterotic/type II string-string
duality for these type of $N=2$ string models,
where we will concentrate on the particular case $N_V=5$. 
Whereas the heterotic moduli $T$ and $U$ are related to the compactification
from six to four dimensions on $T_2$, the additional vector fields
originate from the ten-dimensional gauge group $E_8\times E_8$ which
survive after the compactification on $K3$. Usually the corresponding
complex moduli are called Wilson lines; in case of $N_V=5$ we denote
the single Wilson line vector multiplet by  $V$. The corresponding class
of theories is called $S$-$T$-$U$-$V$ models.

The classical  moduli space as well as the classical $T$-duality
transformations
for heterotic string compactifications with  Wilson line moduli
were derived in \cite{CLM1,COS,MS}. For $p$ non-vanishing Wilson lines
the classical moduli space is locally given by the coset ${SO(2,2+p)\over 
SO(2)\times SO(2+p)}$, and the $T$-duality group is given by 
$SO(2,2+p,{\bf Z})$. Together with the dilaton $S$-field moduli space
one therefore deals at the classical level with the special K\"ahler
spaces ${SU(1,1)\over U(1)}\otimes{SO(2,2+p)\over SO(2)\times SO(2+p)}$,
and the corresponding classical $N=2$ prepotential can be easily
constructed \cite{FP,CDFP,WKLL}. 
At the heterotic one-loop level the effective action is given in terms
of automorphic functions of the duality group $SO(2,2+p,{\bf Z})$, which
are functions of $T$, $U$ and the Wilson line moduli \cite{CLM2,MS}.
One generically encounters singularities at those points in the moduli
space  where certain perturbative BPS states become massless.
(The automorphic functions can be constructed as infinite sum over the
perturbative BPS spectrum.)

In this context it is important to realize that one encounters a very
special situation in the presence of a single Wilson line $V$ only, i.e.
$N_V=5$. In this case, as it was observed in \cite{MS}, the
 classical $T$-duality group $SO(2,3,{\bf Z})$ is isomorphic
to $Sp(4,{\bf Z})$, which has a standard action on the Siegel
upper half plane ${\cal H}_2$.
The corresponding automorphic functions of $Sp(4,{\bf Z})$ are just given
by the Siegel modular forms, which are directly associated
to genus two Riemann surfaces. In the limit of vanishing Wilson line, 
$V\rightarrow 0$, the genus two Riemann surface degenerates into
the product of two $T_2$, and the Siegel modular forms approach
the $SO(2,2,{\bf Z})$ modular functions of the
$S$-$T$-$U$ model in this limit.

In our paper we will show how the heterotic 
one-loop prepotential and the gravitational $F_1$-function for a class of
$N=2$ models with $N_V=5$ can be constructed in terms of 
Siegel modular forms, Jacobi forms and ordinary (functions  of $\tau$ only)
modular functions. The models we are investigating are characterized
by the  embedding of the $SU(2)$ instanton numbers into the
heterotic gauge group $E_8^{(1)}\times E_8^{(2)}$.
We discuss the corresponding dual type II Calabi-Yau compactifications 
with $h_{1,1}=4$ and
find in this way the relation between the relevant modular forms and the
rational
Calabi-Yau instanton numbers. This relation will be shown to be
satisfied for a particular example based on the Calabi-Yau space 
$P_{1,1,2,6,10}(20)$, recently discussed in \cite{BKKM}.

Our paper is organized as follows.
In the next section we define the class of models, that we will be 
investigating
in the following, together  
with their massless spectrum.
The models are discussed from the heterotic as well as from the dual type
II point of view. In particular we discuss the points in the classical
moduli space where extra states become massless. The various
enhancement loci are given in terms
of Humbert surfaces in the
classical moduli space and are related to specific Siegel modular form such as
${\cal C}_{30}(T,U,V)$ and ${\cal C}_5(T,U,V)$.
In section three we present the construction of the supersymmetric index
for the $N=2$ models with one Wilson line. In 3.1. we first review
the computation \cite{HM} of the supersymmetric index of the $S$-$T$-$U$ model.
This construction can be nicely extended to the case $N_V=5$
by a well defined ``hatting" procedure of Jacobi functions, which
describes the transition of going from Jacobi forms to ordinary 
modular forms.
The physical interpretation of the hatting procedure is just the
gauge symmetry breaking $SU(2)\rightarrow U(1)$ 
by turning on the Wilson line $V$.
In section four we use the results of the previous chapter to write
down the heterotic one-loop prepotential as a power series expansion
in terms of hatted Jacobi functions. 
Comparing with the corresponding type II prepotential
we relate the Calabi-Yau instanton numbers to the
coefficients of the heterotic power series expansion. 
Using the known rational instanton numbers for the dual Calabi-Yau
$P_{1,1,2,6,10}(20)$  we show that this relation holds for this
specific example.
In section five we compute the one-loop heterotic function $F_1$
in terms of the Siegel  forms ${\cal C}_{30}$ and ${\cal C}_5$.
A  summary concludes the main body of the paper.
In appendix A we review some interesting
properties of Siegel and Jacobi modular forms. We also provide
more details of the hatting procedure 
and its relation to theta functions and lattices, which is used to construct
the supersymmetric index and the 
heterotic one-loop prepotential
in the presence of a Wilson line $V$. In appendix B we show in some
detail the computation of an integral which is needed for the
computation of $F_1$.

During the process of finishing our calculations and writing
up our results, some related work appeared in \cite{K2}. In \cite{K2}
a four parameter model based on the Calabi-Yau $P_{2,2,3,3,10}(20)$
is discussed. We will make further comments on \cite{K2} in our paper.
It is worth noting that recently 
the Siegel modular forms proved to be relevant for the
computation of the non-perturbative elliptic genus of four-dimensional
$N=4$ strings \cite{DVV}.

\section{$N=2$ four parameter string models}
\setcounter{equation}{0}

In the following, we will discuss a class
of heterotic $4$ parameter 
$N=2$ models, obtained by compactifying
the $E_8 \times E_8$ string on $K3\times T_2$.  The four moduli
comprise the dilaton $S$, the two toroidal moduli $T$ and $U$
as well as a Wilson line $V$.  We will refer to these models
as $S$-$T$-$U$-$V$ models.  
Any of the
$S$-$T$-$U$-$V$ models in the class we will consider here has
a dual type IIA description. Two such duals type II models have been
recently discussed in the literature.  The first one \cite{BKKM}
consists
of a type IIA compactification
on the 
Calabi--Yau three-fold $P_{1,1,2,6,10}(20)$ with
$h_{1,1}=4$, $h_{2,1}=190$ and consequently Euler number $\chi=-372$.
This model has a Higgs transition \cite{BKKM} to the well known
type IIA compactification on $P_{1,1,2,8,12}(24)$ with $h_{1,1}=3$,
$h_{2,1}=243$ and $\chi=-480$, the so-called $S$-$T$-$U$ model \cite{KV}. 
The next 
4 parameter model, discussed by Kawai in \cite{K1,K2}, 
is based on the
Calabi--Yau $P_{2,2,3,3,10}(20)$ with $h_{1,1}=4$, $h_{2,1}=70$ and
$\chi=-132$. 
Finally we will discuss two 4 parameter models with $h_{1,1}=4$, $h_{2,1}=214$,
$\chi =-420$ and $h_{1,1}=4$, $h_{2,1}=202$, $\chi=-396$ respectively;
the corresponding Calabi-Yau spaces were discussed in \cite{CF,LSTY}.
Any of the $S$-$T$-$U$-$V$ models considered here can be 
truncated to the $3$ parameter
$S$-$T$-$U$ model upon setting $V \rightarrow 0$.
Note that this is a truncation as far as 
the vector moduli sector is concerned; in the hyper moduli space
one has to move to a generic point in the course of the
Higgs transition \cite{BKKM}.

The perturbative heterotic $N=2$ models we will consider
in the following will be constructed as follows.  
Following \cite{KV,CF,AFIQ}, we start
with a compactification of the heterotic $E_8^{(1)} \times 
E_8^{(2)}$ string
on $K3$ with $SU(2)$ bundles with instanton numbers $(d_1,d_2)=(12-n,12+n)$.
For $0\leq n\leq 8$, 
the gauge group is $E_7^{(1)}\times E_7^{(2)}$, and the spectrum of
massless hypermultiplets 
follows from the index theorem \cite{GSW,KV} as
\beqa
{1\over 2} (8-n) ({\bf 56},{\bf 1})+{1\over 2}(8+n)({\bf 1},{\bf 56})+
62({\bf 1},{\bf 1}).\label{hmspectrum}
\eeqa
For the standard embedding, $n=12$, 
the gauge group is $E_8^{(1)}\times E_7^{(2)}$
with massless hypermultiplets
\beqa
10({\bf 1},{\bf 56})+65({\bf 1},{\bf 1}).\label{hmspectrum1}
\eeqa
These gauge groups can be further broken by giving vevs to the charged
hypermultiplets. Specifically, $E_7^{(2)}$ can be completely
broken through the chain
\beqa
E_7\rightarrow E_6\rightarrow SO(10)\rightarrow SU(5)\rightarrow
SU(4)\rightarrow SU(3)\rightarrow SU(2)\rightarrow SU(1),\label{chain}
\eeqa
where $SU(1)$ denotes the trivial group consisting of the identity only.
In the following, we will concentrate on the cases where we 
break $E_7^{(2)}$ either completely or down to $SU(2)$.
On the other hand, $E_7^{(1)}$ can be perturbatively 
broken only to some terminal
group $G_0^{(1)}$ that depends on $n$
(see \cite{CF} for details); e.g. for $n=4$ this group is given by $G_0^{(1)}=
SO(8)$.
For $n=8$ it is $G_0^{(1)}=E_8$. It is only for $n=0,1,2$ that $E_7^{(1)}$ 
can be completely 
broken.
Finally, when 
compactifying to four dimensions on $T_2$, three additional vector
fields arise, namely the fields $S$, $T$ and $U$.

Let us first discuss in slightly more detail the class of models where
$E_7^{(2)}$ is completely broken. We will call,
as it will be plausible in the following, these models the ``$S$-$T$-$U$"
class of models.
In the dual type II description the corresponding Calabi-Yau spaces
are given by elliptic fibrations over the Hirzebruch surface $F_n$.
(For $n=2$ the corresponding Calabi-Yau is given by $P_{1,1,2,8,12}(24)$.)
The models with $n=0,1,2$ all contain
$N_V=h_{1,1}+1=4$ Abelian vector multiplets,  the fields $S$, $T$, $U$ plus the
graviphoton, and in addition 
$N_H=h_{2,1}+1=244$  neutral
hypermultiplets. In fact, at the heterotic perturbative 
level all three models are the same; the models with even $n=0,2$ are even
identical at the non-perturbative level.

For  $n>2$ both $N_V$ and $N_H$ increase (see the chain in the first
column of table A.1 in \cite{CF}).
However, suppose that  $G_0^{(1)}$ could be completely
broken and that dim$(G_0^{(1)})$ hypermultiplets
could be made massive by some mechanism, such that the spectrum would be given
by $N_V=4$, $N_H=244$ for all $n$. Then it is natural to conjecture
that 
all models are
perturbatively equivalent;
moreover we conjecture that
the models with even respectively odd $n$ are 
 non-perturbatively
equivalent.

Now let us come to the models with unbroken $SU(2)^{(2)}$. 
The corresponding Hodge numbers are given in the second column
of table A.1 in \cite{CF}.
The universal
vector fields are now given by $S$, $T$, $U$ and $V$, where the
Wilson line $V$ is
in the Cartan subalgebra of $SU(2)^{(2)}$. The commutant of $SU(2)^{(2)}$
in $E_7^{(2)}$ is $SO(12)^{(2)}$.
Then, it follows from the index 
theorem  that the charged spectrum consists of ${1\over 2}(8-n)$
${\bf 56}$ of $E_7^{(1)}$,  as well as of ${1\over 2}(8+n)$ ${\bf 32}$ of
$SO(12)^{(2)}$ plus
62 gauge neutral moduli.

As for the $S$-$T$-$U$ models, it is only possible to perturbatively
 higgs the $E_7^{(1)} \times SO(12)^{(2)}$ completely for $n=0,1,2$.
Thus, these heterotic models 
 will have a massless spectrum comprising $N_V=5$
vector multiplets, $S$, $T$, $U$, $V$ plus the graviphoton,  as well as
\beqa
N_H = ({1\over 2}(8+n) 32 - 66 + 
{1\over 2}(8-n) 56 -133 + 62 = 12 d_1 + 71 = 215 - 12n\label{numberhyper}
\eeqa 
neutral hyper multiplets. Note that, unlike for the $S$-$T$-$U$ models with 
$N_H=244$, the number of hypermultiplets now depends on $n$. Furthermore,
as we will discuss,  for the four parameter models also the
vector multiplet couplings are sensitive to $n$ 
already at the perturbative level.

In the dual type IIA description, 
 based
on compactifications on 4 parameter Calabi--Yau three-folds $X_n$,
the Euler numbers are $\chi(X_n) = 2(h_{1,1} - h_{2,1}) = 24 n -420$
and Hodge numbers are given by $h_{1,1} = N_V- 1 = 4$, $h_{2,1}= N_H-1=
214 -12 n$.  The $n=2$ Calabi--Yau three-fold $X_2$, for instance, is given by
the space $P_{1,1,2,6,10}(20)$ of \cite{BKKM}.
The Calabi--Yau spaces 
$X_0$ and $X_1$ and
%, where no special Higgsing is necessary, 
are given in \cite{CF,LSTY}.

For $n>2$ $E_7^{(1)}$ can only be higgsed to $G_0^{(1)}$
in a perturbative way and hence $N_V>5$.
However, 
suppose again for the moment that $G_0^{(1)}$ can be completely broken
by some mechanism, 
and that dim$(G_0^{(1)})$ massless
hypermultiplets
could disappear. Then $N_V=5$ and the number of massless 
hypermultiplets is given
by eq.(\ref{numberhyper}). This would imply that
on the dual type IIA side there exist Calabi--Yau spaces $X_n$ with 
$\chi(X_n)=24n-420$ for $0\leq 
n\leq8$ and $n=12$. In fact, for $n=12$
a candidate Calabi--Yau really exist, namely the
 $n=12$ Calabi--Yau space $X_{12}$ is given by the space 
$P_{2,2,3,3,10}(20)$ of \cite{K1,K2}.
Note that $X_{12}$ and the $n=2$ space $X_2=P_{1,1,2,6,10}(20)$
both directly show the same K3 fibre $P_{1,1,3,5}(10)$.
Futhermore, this also holds for $X_0$ and $X_1$ \cite{CF}.

In summary, we will focus our proceeding discussion on the cases $n=0,1,2,12$
where the Hodge numbers of the corresponding Calabi-Yau space are summarized 
in the following table.

\hspace{0.5cm}

\begin{center}
\begin{tabular}{|c||c|c|c|c|} \hline
  -      & $X_0$ & $X_1$ & $X_2$ &  $X_{12}$\\ \hline\hline
$-\chi$  &   420 &   396 & 372   &      132   \\ \hline
$h^{2,1}$&   214 &   202 &190    &     70     \\ \hline
\end{tabular}
\end{center}

\hspace{0.5cm}

At the transition point $V=0$, the $U(1)$ associated with the 
Wilson line modulus $V$ becomes enhanced to an $SU(2)$.
Let $N_V'=2 $ and $N_H'$ denote the number of additional
vector and hyper multiplets becoming massless at this transition point.  Then
\beqa
\frac{1}{2} \left( N_H' - N_V' \right) = 6n + 15 \;\;.
\label{nvnh}
\eeqa
This will prove to be a useful relation later on.
It follows from the fact that the Euler number of the Calabi--Yau
space $\chi(X_n)$ and of the $S$-$T$-$U$ models ($\chi = -480$) differ by
$ 2(  N_H' - N_V' ) =\chi(X_n) + 480$.

In addition to the $V=0$ locus of gauge symmetry enhancement, there
are also the enhancement loci (such as $T=U$), associated with the 
toroidal moduli $T$ and $U$, already known from the $S$-$T$-$U$ model.
All these loci correspond to 
surfaces/lines of gauge symmetry enhancement in the
heterotic perturbative moduli space 
${\cal H}_2 = \frac{SO(3,2)}{SO(3)\times SO(2)}$
and have a common description as follows.

Consider the 
Narain lattice $\Gamma=\Lambda\oplus U(-1)$ of signature $(3,2)$, 
where $U(-1)$ denotes the hyperbolic plane
$\left(\begin{array}{cc}0&-1\\-1&0\end{array}\right)$, and where
$\Lambda=U(-1)\oplus<2>=
\left(\begin{array}{ccc}0&-1&0\\-1&0&0\\0&0&2\end{array}\right)$
in a basis which we will denote by
$(f_2,f_{-2},f_3)$; 
we will use the coordinate
$z=iTf_{-2}+iUf_2-iVf_3$ in $\Lambda\otimes {\bf C}$.
Note here that the perturbative moduli space 
$\frac{SO(3,2)}{SO(3)\times SO(2)}$,
which is a hermitian symmetric space, has a representation as a bounded domain
of type IV, that is, as a connected component of 
$ {\cal D}=\{[\omega]\in {\bf P}(\Gamma \otimes {\bf C})|\omega ^2=0,\;
\omega\cdot\bar{\omega}>0\}=\Lambda\otimes {\bf R}+iC(\Lambda)
\subset \Lambda\otimes {\bf C}$, where 
$C(\Lambda)=\{x\in \Lambda\otimes{\bf R}|x^2<0\}$
(this last condition ensures again that 
$2 {\rm Re}T\,{\rm Re} U-2({\rm Re}V)^2>0$; the connected
component can then be realised as 
${\cal D}^+=\Lambda\otimes {\bf R}+iC^+(\Lambda)$
,where $C^+(\Lambda)$ denotes the 
future light cone component of 
$C(\Lambda)$).\\
Now in the basis  
$\varepsilon_1=f_{-2}-f_2,\varepsilon_2=f_3,\varepsilon_3=f_2-f_3$, 
$\Lambda$ is equivalent to the intersection matrix 
$A_{1,0}=\left( \begin{array}{ccc} 2&0&-1\\0&2&-2\\-1&-2&2\end{array}\right)$
associated to the Siegel modular form ${\cal C}_{35}$ of
\cite{GN2}.
To each element $\varepsilon_i$, which squares to $2$, is
associated the 
Weyl reflection $s_i:x\rightarrow x-(x\cdot\varepsilon_i)\varepsilon_i$.
The fixed loci of these Weyl reflections 
give the enhancement loci \cite{CLM2}. As these 
reflection planes are given by planes orthogonal to the elements 
$\varepsilon_i$, 
this gives rise to the following loci: 
the orthogonality conditions $(a\varepsilon_1+b\varepsilon_2+c\varepsilon_3)
\varepsilon_i=0$ yield $c=2a$, $b=c$ and $a=2(c-b)$.  Since $a$, $b$ and
$c$ are related to $T$, $U$ and $V$ by 
$a=iT$, $b=iT+iU-iV$ and $c=iT+iU $, as can be seen 
by comparing 
$a\varepsilon_1+b\varepsilon_2+c\varepsilon_3=a(f_{-2}-f_2)+bf_3+
c(f_2-f_3)=af_{-2}+(c-a)f_2+(b-c)f_3$ 
with $z=iTf_{-2}+iUf_2-iVf_3$, the above orthogonality
conditions result in the enhancement loci $T=U$ ,$V=0$ and $T-2V=0$.
Note that these are 
the conditions
for enhancement loci related to 
${\cal C}_{35}={\cal C}_{30}\cdot{\cal C}_{5}$ (cf. appendix A). 
Also note that the locus
$T - 2V=0$ locus goes over 
into the locus $T-U=0$ 
under the target space duality transformation \cite{CLM1} 
$ T \rightarrow T + U + 2 V, U \rightarrow U, V \rightarrow V + U$.
Thus, the enhancement lines of the $S$-$T$-$U$ 
model have become the Humbert surfaces $H_4$ and $H_1$
(cf. the discussion about 
rational quadratic divisors given in ch. 5 of \cite{HM} ($s=1$)
as well as in 
\cite{GN2}).\\
Furthermore,
\footnote{Note that, since the heterotic perturbative gauge group is 
reflected, 
on the dual
type IIA side, in the (monodromy invariant part of the) Picard group of 
the generic $K3$ fibre of the Calabi-Yau \cite{A,AL}, 
the discussion presented here agrees
precisely with the one of \cite{GN1}
concerning the zero divisor of the period map for the (mirror of the) $K3$.
${\cal D}^+$ can be matched with the domain of the period map $\Phi(z)$.}
for the $K3$-fibre $P_{1,1,3,5}(10)$ of $X_n$,
one finds that (cf. \cite{BKKM} for $n=2$) in the basis $j_1,j_3,j_4$
(where we denote the intersections of the CY divisors with the 
$K3$ ($J_2$) by
small letters) the intersection form is given by 
$\left(\begin{array}{ccc}2&1&4\\1&0&2\\4&2&6\end{array}\right)$, which is
equivalent (over {\bf Z}) to $-\Lambda$ under the base change 
$f_2=j_1-j_3$, $f_{-2}=j_3$ and $f_3=2j_1-j_4$. 
The enhancement loci will become the 
conditions $t_3=0$ resp. $t_4=0$ for the K\"ahler moduli on the type II 
side (cf. section \ref{stumod}).

\section{The supersymmetric index}
\setcounter{equation}{0}

It was shown in \cite{DKL,AGN} that threshold corrections in $N=2$
heterotic string compactifications can be written in terms 
of the supersymmetric index
\beqa
\frac{1}{\eta^2}Tr_R F(-1)^F q^{L_0-c/24}\bar{q}^{\tilde{L}_0-\tilde{c}/24}
\;\;.
\label{susyindex}
\eeqa
This quantity is, as shown in \cite{HM}, also related 
to the computation of the perturbative heterotic $N=2$ prepotential.
In the next subsection we will first review the computation of the 
index (\ref{susyindex}) for an $S$-$T$-$U$ model.  In the following
subsection,  we will then 
discuss its computation in an $S$-$T$-$U$-$V$ model.

\subsection{The $S$-$T$-$U$ models}

For the $S$-$T$-$U$ model with instanton number embedding
$(d_1,d_2)=(0,24)$, the supersymmetric index (\ref{susyindex})
was calculated in \cite{HM} and found to be equal to 
%----------------------------------------------------------
\begin{eqnarray}
\frac{1}{\eta^2}Tr_R F(-1)^F q^{L_0-c/24}\bar{q}^{\tilde{L}_0-\tilde{c}/24}
=-2iZ_{2,2}\frac{E_4E_6}{\Delta} \;\;,
\label{sindhm}
\end{eqnarray}
%----------------------------------------------------------
where $Z_{2,2}$ denotes the sum over the Narain lattice $\Gamma_{2,2}$,
$Z_{2,2}=\sum_{p\in \Gamma^{2,2}}q^\frac{p_L^2}{2}\bar{q}^\frac{p_R^2}{2}$,
and where
$\frac{E_4E_6}{\Delta} = \sum_{n\ge -1}\tilde c_{STU}(n)q^n$. 
Here the subscript on 
the trace indicates the Ramond sector as right-moving boundary condition;
$F$ denotes the right-moving fermion number, $F=F_R$.\\
Let us recall how this expression came about. First, one can reduce 
(\ref{susyindex}) to 
$\frac{1}{\eta^2}Tr_R (-1)^F q^{L_0-c/24}\bar{q}^{\tilde{L}_0-\tilde{c}/24}$,
where the contributions are weighted with $\pm 2\pi i$ depending on
whether a BPS hyper or vector multiplet contributes.
The expression resulting from the evaluation of the trace consists of 
the product of three terms, namely of 
$Z_{2,2}/\eta^4$, of the partition function for the first 
$E_8^{(1)}$ in the bosonic 
formulation (leading to the contribution $E_4/\eta^8$)
and of the elliptic genus for the second $E_8^{(2)}$ 
containing the gauge connection on $K3$.\\
This last quantity decomposes now additively (taking into account the 
appropriate weightings) into contributions from the
following sectors, namely: 1) the 
$(NS,R)$ sector, which we will also denote by $(NS^+,R)$, 
2) the ``twisted" sector
$(NS^-,R)$, where a factor $(-1)^{F_L}$ is inserted in the trace 
(this contribution is weighted with (-1)) 
and 3) the $(R,R)$ sector, which we will also denote by $(R^+,R)$.
Since we are using the fermionic representation
for $E_8^{(2)}$, we decompose
the fermionic $D_8^{(2)}\subset E_8^{(2)}$, so that
each of these contributions splits again multiplicatively
into a 
free $D_6^{(2)}$ part and into a $D_2$ part,
 to be called $D_2^{(2)}K3$, 
containing the gauge connection $A_1$ which describes the corresponding
gauge bundle on $K3$.  The corresponding contributions are 
summarized in the following table, where we also indicate the connection
to the generic elliptic genus 
$Z(\tau,z)=
Tr_{R,R} y^{F_L}(-1)^{F_L+F_R}q^{L_o-c/24}\bar{q}^{\tilde{L}_o-\tilde{c}/24}=
6\frac{\th_2^2\th_4^2}{\eta^4}\frac{\th_3^2(\tau,z)}{\eta^2}
-2\frac{\th_4^4-\th_2^4}{\eta^4}\frac{\th_1^2(\tau,z)}{\eta^2}$,
where
$y={\bf e}[z]={\rm exp} 2\pi i z$  (cf. \cite{EOTY,KYY}).

\hspace {0.5cm}

\begin{center}
\begin{tabular}{|c||c|c|}\hline
Tr&$D_6$&$K3D_2$\\\hline\hline
$(NS^+,R)$&$\frac{\th_3^6}{\eta^6}$
           &$-2\frac{\th_4^4-\th_2^4}{\eta^4}\frac{\th_3^2}{\eta^2}
           =q^{\frac{1}{4}}Z(\tau,\frac{\tau +1}{2})$\\
$(NS^-,R)$&$\frac{\th_4^6}{\eta^6}$
           &$2\frac{\th_2^4+\th_3^4}{\eta^4}\frac{\th_4^2}{\eta^2}
           =q^{\frac{1}{4}}Z(\tau,\frac{\tau}{2})$\\
$(R^+,R)$ &$\frac{\th_2^6}{\eta^6}$           
           &$2\frac{\th_3^4+\th_4^4}{\eta^4}\frac{\th_2^2}{\eta^2}
           =Z(\tau,\frac{1}{2})$\\
$(R^-,R)$ &$\frac{\th_1^6}{\eta^6}=0$
           &$6\frac{\th_2^2\th_3^2\th_4^2}{\eta^4\cdot\eta^2}=24
           =Z(\tau,0)$\\ \hline
\end{tabular}
\end{center}

\hspace {0.5cm}

Now recall
that $E_4$ and $E_6$ have the following $\theta$-function decomposition
\beqa
2 E_4
&=&\th_2^6\cdot\th_2^2 +\th_3^6\cdot\th_3^2 +
\th_4^6\cdot\th_4^2  \nonumber\\
2 E_6&=&-\th_2^6(\th_3^4+\th_4^4)\cdot\th_2^2+
\th_3^6(\th_4^4-\th_2^4)\cdot\th_3^2+
\th_4^6(\th_2^4+\th_3^4)\cdot\th_4^2  \;\; ;
\label{e4e6}
\eeqa
the $\theta_i^2$ 
contributions ($i=2,3,4$) are due to the $SO(4)$ piece in the 
fermionic decomposition of $E_8 \supset SO(12) \times SO(4)$.
Hence the sum of the three non-vanishing terms in the table 
precisely leads to (\ref{sindhm}).

On the other hand, in the case of a general $(d_1,d_2)$ embedding 
(using now a fermionic representation for both $E_8$'s),
one first has to decompose the $D_2^{(1)}K3D_2^{(2)}$ part
 into 
$D_2^{(1)}K3\times D_2^{{(2)},free}+D_2^{{(1)},free}\times K3D_2^{(2)}$, 
where the factors in each summand
are now in different, and hence commuting, $E_8$'s.  Furthermore, since
the rudimentary $K3$ gauge bundles are structurally completely the same
as before, the amount of contribution realised by them can - by 
comparison with the ``complete" $K3$ bundle considered above - be read off 
from the $R^-$ sector.  Note that $Z(\tau,0)$ is the Witten
index, which gives the Euler number of $K3$ resp. the second Chern class of
the relevant vector bundle. \\
This results in a contribution proportional to 
%----------------------------------------------------------
\begin{eqnarray}
\frac{1}{\Delta}(\frac{d_1}{24}E_6\cdot E_4+E_4\cdot \frac{d_2}{24}E_6)
=\frac{1}{\Delta}E_4E_6 
=\sum_{n\ge -1} c_{STU}(n)q^n
\;\;,
\end{eqnarray}
%----------------------------------------------------------
so that the result is independent of the particular instanton embedding.
Related universality properties of $N=2$ threshold corrections
in case of vanishing Wilson lines were
discussed in \cite{Kir}.

\subsection{The $S$-$T$-$U$-$V$ models}

In the presence of a Wilson line, which we will take to lay in the
second $E_8^{(2)}$, 
the symmetry between the two $E_8$'s is broken and thus,
contrary to the 3 parameter case,  the prepotential
will already depend perturbatively on the type $(d_1,d_2)$ of the 
instanton embedding (we take $d_2\ge d_1$).\\
The supersymmetric index (\ref{susyindex}) will now have the form
\beqa
\frac{1}{\eta^2}Tr_R F(-1)^F q^{L_0-c/24}\bar{q}^{\tilde{L}_0-\tilde{c}/24}
=-2iZ_{3,2}(\tau, {\bar \tau}) F(\tau) \;\;,
\label{indexstuv}
\eeqa
where $Z_{3,2}$ denotes the sum over the Narain lattice $\Gamma_{3,2}$,
$Z_{3,2}=\sum_{p\in \Gamma^{3,2}}q^\frac{p_L^2}{2}\bar{q}^\frac{p_R^2}{2}$.
The presence of the Wilson line in $E_8^{(2)}$
has the following effect on the $\theta_i^2$ pieces
appearing in the decomposition (\ref{e4e6}) of $E_4$ and $E_6$
\beqa
2 \widehat{E_{4,1}(\tau,z)}
&=&\th_2^6\cdot \widehat{\th_2^2(\tau,z)} +\th_3^6\cdot 
\widehat{\th_3^2(\tau,z)}+
\th_4^6\cdot \widehat{\th_4^2(\tau,z)}  \;\;,  \\
2 \widehat{E_{6,1}(\tau,z)
}&=&-\th_2^6(\th_3^4+\th_4^4)\cdot\widehat{\th_2^2(\tau,z)}
+ \th_3^6(\th_4^4-\th_2^4)\cdot \widehat{\th_3^2(\tau,z)}+
\th_4^6(\th_2^4+\th_3^4)\cdot  \widehat{\th_4^2(\tau,z)} \;\;,\nonumber
\eeqa
where
\begin{eqnarray}
\widehat{\th_1^2(\tau,z)}&=&\th_2(2\tau)-\th_3(2\tau)\;\;, \nonumber\\
\widehat{\th_2^2(\tau,z)}&=&\th_2(2\tau)+\th_3(2\tau)\;\;,\nonumber\\
\widehat{\th_3^2(\tau,z)}&=&\th_3(2\tau)+\th_2(2\tau)\;\;,\nonumber\\
\widehat{\th_4^2(\tau,z)}&=&\th_3(2\tau)-\th_2(2\tau) 
\end{eqnarray}
%------------------------------------------------------------
are the two $SU(2)$ characters of the surviving $A_1$ when written in
the boundary condition picture instead of the usual conjugacy class picture.
We refer to appendix \ref{thefjaf} and A.5 for description and interpretation
of the hatting procedure.

The replacement $E_4  \rightarrow \widehat{E_{4,1}}$, in particular,
amounts to replacing the $E_8$ partition function 
$P_{E_8}=P_{E_7^{(0)}}\cdot P_{A_1^{(0)}}+P_{E_7^{(1)}}\cdot P_{A_1^{(1)}}$
with $P_{E_7^{(0)}}+P_{E_7^{(1)}}$.  This precisely describes the breaking of
the $E_8^{(2)}$ to $E_7^{(2)}\times U(1)$ when turning on a Wilson line.

Thus, the effect of turning on a Wilson line can be described as follows.
Introducing 
%----------------------------------------------------------
\begin{eqnarray}
A_n(\tau)=\frac{1}{\Delta} \left(\frac{d_1}{24}E_6\cdot\widehat{E_{4,1}}+
E_4\cdot\frac{d_2}{24}\widehat{E_{6,1}}\right) \;\;,
\label{An}
\end{eqnarray}
%----------------------------------------------------------
it follows that turning on a Wilson line results in the replacement
\beqa
Z_{2,2} &\rightarrow & Z_{3,2}\;\;, \nonumber\\
\frac{1}{\Delta} \left(\frac{d_1}{24}E_6\cdot E_4
+ E_4\cdot \frac{d_2}{24}E_6\right) & \rightarrow &
F(\tau) = A_n \;\;.
\eeqa
(The first few expansion coefficients of $A_0$, $A_1$, $A_2$ and $A_{12}$
are listed in the second table in appendix A.6.)
The product
$Z_{3,2} A_n$ transforms covariantly under modular transformations, since
$F(\tau)$ has weight $-2\frac{1}{2}$. (Recall that ${E_4E_6\over \Delta}$
has weight -2.)

The occurence of modular forms $F(\tau)$ 
of half-integral weight is naturally 
understood by realising that 
the present case (of $s=1$ Wilson lines turned on)
interpolates between the $s=0$ and $s=8$ cases
of \cite{HM}, where the relevant modular forms
$E_4E_6/\Delta$ ($s=0$) 
and $E_6/\Delta$ ($s=8$) are of weight $-2$ and $-6$, respectively.

\section{The perturbative prepotential for the $S$-$T$-$U$-$V$ models
\label{stumod}}
\setcounter{equation}{0}

In this section we discuss the 
relation
between the type II and the heterotic prepotentials for the 
$S$-$T$-$U$-$V$ models, that is between
rational instanton numbers on the type II side and 
Siegel modular forms on the heterotic side. 
The appearance
of Siegel modular forms in the context of threshold corrections in the
presence of Wilson lines 
was first pointed out in \cite{MS}.

As discussed in the previous section, the supersymmetric index
is given in terms of 
\beqa
F(\tau) = A_n = \sum_{N \in  {\bf Z}, {\bf Z} + \frac{3}{4}}
c_n(4N) q^N \;\;.
\eeqa
As explained in appendix A, the modular function 
$A_n(\tau)$ is in one-to-one correspondence with the index-one 
Jacobi form with the same expansion coefficients $c_n(k,b)=c_n(4k-b^2)$:
$A_n(\tau)=\widehat{A_n(\tau,z)}$, $A_n(\tau,z)={1\over \Delta(\tau)}
\left(\frac{d_1}{24}E_6(\tau)\cdot E_{4,1}(\tau,z)+
E_4(\tau)\cdot\frac{d_2}{24}E_{6,1}(\tau,z)\right)=
\sum_{k,b}c_n(4k-b^2)q^kr^b$.\footnote{$A_n(\tau,z)$
can be eventually seen as the order $s$ expansion coefficient
of a Siegel modular form $F_n(T,U,V)$, again with identical expansion
coefficients. Specifically, the index-one Jacobi  form $A_n(\tau,z)$
is the order $s$ expansion coefficient of the Siegel form
${1\over 132}
(-
{\cal E}_4
{\cal E}_6+(
31\cdot 12^3-11\cdot 12^2 n){\cal C}_{10})$.}

The expansion coefficients
 $c_n(4N)$ of $F(\tau)$  govern
the perturbative, i.e. 1-loop,
 corrections to the heterotic prepotential $F_0^{\rm het}$
\cite{HM}.   
For the class of $S$-$T$-$U$-$V$ models considered
here, the perturbative
heterotic prepotential is given by
\beqa
F_0^{\rm het} &=& - S (TU - V^2) + p_n(T,U,V) - 
\frac{1}{4\pi^3}\sum_{k,l,b\in {\bf Z} \atop
(k,l,b)>0} c_n(4kl-b^2) Li_3({\bf e}[kiT+liU+biV]), 
\nonumber\\
\label{prepstuv}
\eeqa
where 
${\bf e}[x]={\rm exp} 2 \pi i x$.
The first term $ - S(TU-V^2)$ is the tree-level
prepotential of the 
special 
K\"ahler space ${SO(3,2)\over SO(3)\times SO(2)}$;
 $p_n(T,U,V)$ denotes the one-loop cubic polynomial which depends on the
particular instanton embedding $n$.
The
condition
$(k,l,b)>0$ means that: either $k > 0, l , b \in {\bf Z}$ or 
$k=0,l>0, b \in {\bf Z}$ 
or $k=l=0, b <0$ (cf. \cite{HM}).
It is shown in appendix \ref{wsint} how the worldsheet expansion coefficients
$c_n(4N)$ turn into the target-space coefficients $c_n(4kl-b^2)$ appearing
in the prepotential. 

Next, consider truncating an $S$-$T$-$U$-$V$ model
to the $S$-$T$-$U$ model by setting $V=0$.  Then, the sum over $b$
in (\ref{prepstuv}) yields independently from $n$ the coefficients of the 
$3$ parameter model,
\beqa
c_{STU}(kl)=\sum_b c_n(4kl-b^2)  \;\;,
\eeqa
as it can be checked by explicit comparison.
Therefore the prepotential (\ref{prepstuv}) truncates correctly to the 
prepotential for the $S$-$T$-$U$ model.

The (Wilsonian) 
Abelian gauge threshold functions are related (see \cite{WKLL} for details)
to the second derivatives
of the one-loop prepotential 
$h(T,U,V) = p_n(T,U,V) - \frac{1}{4 \pi^3} \sum_{(k,l,b)>0} c_n(4kl-b^2)
Li_3({\bf e}[kiT+liU+biV])$. 
At the loci of enhanced non-Abelian gauge symmetries some of the
Abelian
gauge couplings will exhibit logarithmic singularities due
to the additional massless states.
First consider $\partial_T\partial_U h$.
At the line $T=U$  one $U(1)$ is extended to $SU(2)$ without
additional massless hypermultiplets. 
It can  be easily checked that, as $T \rightarrow U$,
\beqa
\partial_T \partial_U h = -\frac{1}{\pi} \log(T-U) \;\;,
\eeqa
as it should.
The Siegel modular form which vanishes on the $T=U$ locus and has
modular weight $0$ is given by $\frac{{\cal C}^2_{30}}
{{\cal C}_{12}^5}$.
It can be shown that, as $V \rightarrow 0$,
\beqa
 \frac{{\cal C}^2_{30}}{{\cal C}_{12}^5} \rightarrow (j(T) - j(U))^2\;\;,
\eeqa
up to a normalization constant.
Hence one deduces that
\beqa
\partial_T \partial_U h =
-\frac{1}{2\pi} \log \frac{{\cal C}^2_{30}}{{\cal C}_{12}^5}+ regular.
\label{tugaugecoupl}
\eeqa
On the other hand, at the locus $V=0$, a different $U(1)$ gets enhanced to
$SU(2)^{(2)}$, and at the same time $N_H'$ hyper multiplets, 
being doublets of $SU(2)^{(2)}$,
become massless. Using eq.(\ref{nvnh}), $N_V'=2$ and that
$c_n(-1)=-N_H',c_n(-4)=N_V'$, it can be checked that, as
 $V\rightarrow 0$,
\beqa
-{1\over 4}\partial_V^2 h  =
\frac{3}{2\pi} (2 + n) \log V=
-{1\over \pi}
(1-{1\over 8}N_H')\log V \;\;.
\eeqa
Observe that the factor $(1-{1\over 8}N_H')$ is precisely given
by the $N=2$ $SU(2)$ gauge $\beta$-function coefficient with 
$N_H'/2$ hypermultiplets in the fundamental representation of $SU(2)$.
The Siegel modular form which vanishes on the $V=0$ locus and has
modular weight $0$ is given by $\frac{{\cal C}_5}
{{\cal C}_{12}^{5/12}}$.
It can be shown that, as $V \rightarrow 0$, 
\beqa
{\cal C}_5 \rightarrow V \left(\Delta(T) \Delta(U) \right)^{\frac{1}{2}}
\;\;,
\eeqa
So we now conclude that
\beqa
-{1\over 4}\partial_V^2 h = \frac{3}{4\pi} (2 + n) \log \left(\frac{{\cal C}_5}
{{\cal C}_{12}^{5/12}}\right)^2+ regular.\label{vgaugecoupl}
\eeqa

Let us now compare the heterotic models with the corresponding type
II models on the Calabi--Yau spaces $X_n$. 
The cubic parts of
the type II prepotentials of $X_0$, $X_1$ and $X_2$
are given in \cite{BKKM,LSTY} and can be written in an universal,
$n$-dependent function as follows:
\beqa
F^{II}_{\rm cubic}&=&t_2(t_1^2+t_1t_3+4t_1t_4+2t_3t_4+3t_4^2)\nonumber \\
&+&{4\over 3}t_1^3+8t_1^2t_4+{n\over 2}t_1t_3^2
+(1+{n\over 2})t_1^2t_3 +2(n+2)t_1t_3t_4\nonumber \\
&+&nt_3^2t_4
+(14-n)t_1t_4^2 
+ (4+n)t_3t_4^2+(8-n)t_4^3 . \label{cubicf}
\eeqa
We believe that this expression is also valid for $X_n$ 
with $n>2$, in particular
also for the Calabi--Yau model $X_{12}$.
Note that for $t_4=0$, $F^{II}_{\rm cubic}$ 
precisely reduces to the cubic prepotential
of the $S$-$T$-$U$ models \cite{HKTY,LSTY}. In order to match (\ref{cubicf})
with the cubic part of the heterotic prepotential given in 
(\ref{prepstuv}), we will perform the following identification of
type II and heterotic moduli (which differs from the one 
given in \cite{BKKM})
\beqa
t_1&=&U-2V,\qquad t_2= S-{n\over 2}T-(1-{n\over 2})U,\nonumber\\
t_3&=&T-U,\qquad t_4=V  \;\;,
\label{coordin}
\eeqa
which is valid in the chamber $T>U>2V$.  
Then, (\ref{cubicf}) turns into
\beqa
F^{II}_{\rm cubic}= - F^{\rm het}_{\rm cubic} = S(TU-V^2)
+{1\over 3}U^3+({4\over 3}+n)V^3-(1+{n\over 2})UV^2-{n\over 2}
TV^2 \;\;.\label{cubicfa}
\eeqa
Note that using the heterotic moduli the prepotential is independent of $n$
in the limit $V=0$.

Next, let us consider the contributions of the world sheet 
instantons to the type II prepotential of a 4 parameter model.
  Generically, they are given by 
\beqa
F^{II}_{\rm inst} =-{1\over ( 2\pi)^3}\sum_{d_1,\dots , d_4}n^r_{d_1,\dots ,
d_4}Li_3(\prod_{i=1}^4q^{d_i}) \;\;.
\label{instanton}
\eeqa
The $n^r_{d_1,d_2,d_3,d_4}$ denote the rational instanton numbers.
The heterotic weak coupling limit $S \rightarrow \infty$ corresponds
to the large K\"ahler class limit $t_2 \rightarrow \infty$.  In this
limit, only the instanton numbers with $d_2=0$ contribute
in the above sum.  Using the identification 
$kT+lU+bV=d_1t_1+d_3t_3+d_4t_4$, it follows that (independently of $n$)
\begin{eqnarray}
k&=&d_3 \;\;, \nonumber\\
l&=&d_1-d_3   \;\;,\nonumber\\
b&=&d_4-2d_1  \;\;.
\end{eqnarray}
Then, (\ref{instanton}) turns into
\beqa
F^{II}_{\rm inst}=-{1\over (2\pi)^3}\sum_{k,l,b}n^r_{k,l,b}
Li_3(e^{-2\pi(kT+lU+bV)}) \;\;.
\label{largespr}
\eeqa
Comparison with (\ref{prepstuv}) shows that the rational instanton
numbers have to satisfy the following constraint
\beqa
n^r_{k,l,b}=n^r(4kl-b^2)\label{constraint}
\eeqa
as well as
\beqa
n^r_{k,l,b} = - 2 c_n(4kl-b^2) \;\;.
\label{insmod}
\eeqa
Note that 
the constraint (\ref{constraint}) is non-trivial.  We 
conjecture that an analogous
constraint has to hold for an arbitrary number of Wilson lines
after the proper identification of $T$ and $U$. 
Also note that $c_n(0)=\chi(X_n)$ and 
racall that $c_n(-1)=-N_H',c_n(-4)=N_V'$ .

For concreteness, let us now check above relations 
for the $4$ parameter model of
\cite{BKKM}, which has a dual type II description based on the
Calabi--Yau space $X_2=P_{1,1,2,6,10}(20)$. 
Using the instanton numbers given in \cite{BKKM}\footnote{We are
grateful to B. Andreas and P. Mayr for providing us the
higher instanton numbers which are not given in \cite{BKKM}.}, it can be
checked that both (\ref{constraint}) and (\ref{insmod}) for $c_2$
indeed hold,
 as can be seen from the second table in appendix A.6 and the table given below.

\hspace{0.5cm}

\begin{center}

\begin{tabular}{|ccc|ccc||c|c|} \hline
$d_1$ & $d_3$ & $d_4$ & k & l & b & $N=4kl-b^2$ &$n_{d_1,0,d_3,d_4}$\\ 
\hline\hline
  0   &   0   &   3   & 0 & 0 & 3 &      -9     &          0        \\ \hline
  0   &   1   &   0   & 1 & -1 & 0 &      -4     &         -2        \\
  0   &   0   &   2   & 0 & 0 & 2 &      -4     &         -2        \\
  1   &   0   &   0   & 0 & 1 &-2 &      -4     &         -2        \\
  1   &   0   &   4   & 0 & 1 &2 &      -4     &         -2        \\  \hline 
  0   &   0   &   1   & 0 & 0 & 1 &      -1     &         56        \\
  1   &   0   &   3   & 0 & 1 & 1 &      -1     &         56        \\
  1   &   0   &   1   & 0 & 1 &-1 &      -1     &         56        \\  \hline
  1   &   0   &   2   & 0 & 1 & 0 &       0     &        372        \\  \hline
  2   &   1   &   3   & 1 & 1 &-1 &       3     &      53952        \\  \hline
  2   &   1   &   4   & 1 & 1 & 0 &       4     &     174240        \\  \hline
\end{tabular}

\end{center}

\hspace{0.5cm}

The truncation to the three parameter
Calabi--Yau model is made by setting $V=0$.
The instanton numbers $n^{r}_{k,l}$ of the $S$-$T$-$U$ model are then 
given by \cite{BKKM}
\beqa
n^{r}_{k,l} = 
\sum_{b} n^r(4kl-b^2) \;\;,\label{truncsum}
\eeqa
where the summation range over $b$ is finite.  
For example, $n^r_{1,0}= -2 + 56 + 372 +56 -2 = 480$
\cite{BKKM}.\\

\section{The heterotic perturbative Wilsonian gravitational coupling $F_1$}
\setcounter{equation}{0}
\subsection{BPS orbits}

An important role in the computation of the Wilsonian
gravitational coupling $F_1$ is played by BPS states \cite{FKLZ,VAFA,HM,CCLMR},
\beqa
F_1 \propto \log {\cal M}  \;\;,
\eeqa
where ${\cal M}$ denotes the moduli-dependent holomorphic mass
of an $N=2$ BPS state.  For the $S$-$T$-$U$-$V$ models
under consideration, the tree-level mass ${\cal M}$ is given by
\cite{CLM2,K1,NEU}
\beqa
{\cal M} = m_2 - i m_1 U + in_1 T + n_2 (-UT + V^2) + i b V  \;\;.
\eeqa 
Here, $l=(n_1,m_1,n_2,m_2,b)$ denotes the set of integral quantum numbers 
carried by the BPS state. 
The level matching condition for 
a BPS state reads 
\beqa
2(p^2_L - p^2_R) = 4 n^Tm + b^2  \;\;.
\eeqa
Of special relevance to the computation of perturbative
corrections to $F_1$ are those BPS states, whose tree-level mass vanishes
at certain surfaces/lines in the perturbative
moduli space ${\cal H}_2=\frac{SO(3,2)}{SO(3)\times SO(2)}$.
Note that the condition ${\cal M}=0$ is the
condition (see appendix \ref{hump}) for a rational quadratic 
divisor 
\beqa
{\sc H}_l=\{\left( \begin{array}{cc}i T &i V \\i V &i U \end{array} \right)
\in {\cal H}_2| m_2 - i m_1 U + in_1 T + n_2 (-UT + V^2) + i b V =0 \}
\eeqa 
of discriminant
\beqa
{\rm D}(l)= 2(p^2_L - p^2_R) = 
4m_1n_1+4n_2m_2 + b^2 \;\;.
\eeqa
Consider, for instance, BPS states becoming massless at the surface $V=0$,
the so-called Humbert surface $H_{\rm 1}$ (cf. appendix \ref{hump}).
They lay on the orbit ${\rm D}(l)=1$, that is, on the orbit $n^Tm=0, 
b^2=1$.  On the other hand, BPS states becoming massless
at $T=U$, the Humbert surface $H_{\rm 4}$, lay on the orbit ${\rm D}(l)=4$,
that is, they carry quantum numbers satisfying $n^T m=1, b^2=0$ \cite{CLM2}.

\subsection{The coupling $F_1$ in the $S$-$T$-$U$ model}

The perturbative Wilsonian gravitational coupling for the $S$-$T$-$U$ model
is given by\footnote{The dilaton is defined to be $S=4 \pi/g^2 - 
i \theta/2 \pi$.} (in the chamber $T > U$)
\beqa
F_1 &=& 24 S_{\rm inv} - \frac{b_{\rm grav}}{\pi} \log \eta(T) \eta(U)
+ \frac{2}{\pi} \log (j(T)-j(U)) \;\;.
\label{f1stu}
\eeqa
Using that \cite{WKLL}
\beqa
S_{\rm inv} &=& {\tilde S} + \frac{1}{8} L \;\;, \nonumber\\
{\tilde S} &=& S - \frac{1}{2} \partial_T \partial_U h \;\;\; , \;\;\;
L=-\frac{4}{ \pi} \log(j(T)-j(U)) \;\;,
\eeqa
it follows that $F_1$ can be rewritten as 
\beqa
F_1 = 24 \tilde{S} - \frac{1}{\pi} \Big[10 \log (j(T) - j(U))
+ b_{\rm grav} \log \eta(T) \eta(U) \Big] \;\;.
\eeqa
The perturbative 
gravitational coupling is related to the perturbative Wilsonian coupling
by
\beqa
\frac{1}{g^2_{\rm grav}} &=& \Re F_1 + \frac{b_{\rm grav}}{4 \pi} K 
= 12(S + {\bar S} + V_{GS}) + \Delta_{\rm grav} \;\;.
\label{fgrav}
\eeqa
This relates the Wilsonian gravitational coupling $F_1$ to the
supersymmetric index, that is to 
$\Delta_{\rm grav}= - \frac{2}{4 \pi} {\tilde I}_{2,2}$ \cite{CCLMR},
where \cite{HM}
\beqa
{\tilde I}_{2,2} = \int_{\cal F} \frac{d^2 \tau}{\tau_2}
\Big[ Z_{2,2} \frac{E_4E_6}{\eta^{24}} \big(E_2 - \frac{3}{\pi \tau_2}\Big)
- {\tilde c}_1(0) \Big]
\;\;.
\eeqa
It follows from (\ref{fgrav}) that 
\beqa
F_1 &=& 24 S - \frac{2}{\pi} \sum_{r >0} \tilde{c}_1 (-\frac{r^2}{2}) Li_1
\nonumber\\
&=& 24 \tilde{S} - \frac{1}{\pi} 
 \Big[10 \log (j(T) - j(U))
+ b_{\rm grav} \log \eta(T) \eta(U) \Big] \;\;,
\label{stuf1}
\eeqa
where the coefficients $\tilde{c}_1$ are given by \cite{HM}
\beqa
\frac{E_2E_4E_6}{\Delta} = \sum \tilde{c}_1(n) q^n \;\;,\;\; 
\Delta = \eta^{24} \;\;.
\eeqa
Here, we have ignored the issue of ambiguities in (\ref{stuf1})
linear in $T$ and in $U$.

\subsection{The coupling $F_1$ in the $S$-$T$-$U$-$V$ models}

The classical moduli space of a heterotic $S$-$T$-$U$-$V$ model
is locally given by the Siegel upper half plane ${\cal H}_2 = 
\frac{SO(3,2)}{SO(3) \times SO(2)}$. Because of target space 
duality invariance, 
one has to consider modular forms on ${\cal H}_2$, i.e.
Siegel modular forms (cf. appendix A).

The Siegel modular form which vanishes on the $T=U$ locus and has
modular weight $0$ is given by $\frac{{\cal C}^2_{30}}
{{\cal C}_{12}^5}$.
It can be shown that, as $V \rightarrow 0$,
\beqa
 \frac{{\cal C}^2_{30}}{{\cal C}_{12}^5} \rightarrow (j(T) - j(U))^2\;\;,
\eeqa
up to a normalization constant.
On the other hand, 
the Siegel modular form which vanishes on the $V=0$ locus and has
modular weight $0$ is given by $\frac{{\cal C}_5}
{{\cal C}_{12}^{5/12}}$.
It can be shown that, as $V \rightarrow 0$, 
\beqa
{\cal C}_5 \rightarrow V \left(\Delta(T) \Delta(U) \right)^{\frac{1}{2}}
\;\;,
\eeqa
up to a proportionality constant.   
Finally, the Siegel form ${\cal C}_{12}$ generalises $\Delta(T) \Delta(U)$,
that is
\beqa
{\cal C}_{12} \rightarrow \Delta(T) \Delta(U)
\eeqa 
as $V \rightarrow 0$. \\
Then,
in analogy to (\ref{f1stu}), the perturbative Wilsonian gravitational
coupling for an $S$-$T$-$U$-$V$ model is now given by (in the 
chamber $T> U$)
\beqa
F_1 &=& 24 S_{\rm inv} - \frac{b_{\rm grav}}{24 \pi} \log {\cal C}_{12}
+ \frac{1}{\pi} \log \frac{{\cal C}^2_{30}}{{\cal C}_{12}^5} 
- \frac{1}{2 \pi} (N_H'-N_V') \log \left(
\frac{{\cal C}_5}{{\cal C}_{12}^{5/12}}\right)^2 \;\;.
\label{f1stuv}
\eeqa
Here, $N_V'$ and $N_H'$ denote the vector and the hyper multiplets
which become massless at the $V=0$ locus. Since at $V=0$ there is 
a gauge symmetry restoration $U(1) \rightarrow SU(2)$, we have
$N_V'=2$.\\
The invariant dilaton $S_{\rm inv}$ is given by \cite{WKLL}
\beqa
S_{\rm inv} &=& {\tilde S} + \frac{1}{10} L \;\;, \nonumber\\
{\tilde S} &=& S - \frac{4}{10}(\partial_T \partial_U - 
\frac{1}{4} \partial_V^2) h \;\;,
\eeqa
where the role of the quantity $L$ is to render $S_{\rm inv}$ free
of singularities.
Using eqs.(\ref{tugaugecoupl}) and (\ref{vgaugecoupl}),
it follows that
\beqa
{\tilde S} = S + \frac{1}{5 \pi} \log \frac{{\cal C}^2_{30}}
{{\cal C}_{12}^5} 
- \frac{3}{10 \pi} (2 + n) \log \left(
\frac{{\cal C}_5}{{\cal C}_{12}^{5/12}}\right)^2
+ regular
\eeqa
and, hence,
\beqa
L=-\frac{2}{\pi} \log \frac{{\cal C}^2_{30}}{{\cal C}_{12}^5} 
+ \frac{3}{ \pi} (2 + n) \log \left(\frac{{\cal C}_5}
{{\cal C}_{12}^{5/12}}\right)^2 \;\;.
\eeqa
It follows that the Wilsonian gravitational coupling (\ref{f1stuv})
can be rewritten into
\beqa
F_1 &=& 24 {\tilde S} - \frac{1}{\pi} \Big[
\frac{19}{5} \log \frac{{\cal C}^2_{30}}{{\cal C}_{12}^5} + 
\frac{b_{\rm grav}}{24} \log {\cal C}_{12} \nonumber\\
&+& 
\left( - \frac{72}{10}(2+n) + \frac{1}{2}(N_H'-N_V') \right)
 \log \left(\frac{{\cal C}_5}{{\cal C}_{12}^{5/12}}\right)^2 \Big] \;\;.
\label{f1tildev}
\eeqa
Now recall from (\ref{nvnh}) that $N_H'-N_V'= 12n + 30$.  Inserting 
this into (\ref{f1tildev}) yields 
\beqa
F_1 &=& 24 {\tilde S} - \frac{1}{\pi} \Big[
\frac{19}{5} \log {\cal C}^2_{30} +  
  \frac{3}{5}(1-2n) 
 \log {\cal C}_5^2 \Big] \;\;.
\label{f1tildestuv}
\eeqa
Note that the $\log {\cal C}_{12}$ terms have completely canceled out!

Now consider the perturbative 
gravitational coupling, which is again related to 
the perturbative Wilsonian coupling by
\beqa
\frac{1}{g^2_{\rm grav}} &=& \Re F_1 + \frac{b_{\rm grav}}{4 \pi} K 
= 12(S + {\bar S} + V_{GS}) + \Delta_{\rm grav} \;\;,
\label{fgravv}
\eeqa
where this time
$\Delta_{\rm grav}= - \frac{2}{4 \pi} {\tilde I}_{3,2}$ with
\beqa
{\tilde I}_{3,2} = \int_{\cal F} \frac{d^2 \tau}{\tau_2}
\Big[ Z_{3,2} A_n \Big( E_2 - \frac{3}{\pi \tau_2}\Big)
- d_n(0) \Big]
\;\;.
\label{intws}
\eeqa
Here, we have introduced
\beqa
B_n(\tau) &=& E_2 A_n = \frac{12-n}{24} 
\frac{E_2 E_6 \widehat{ E_{4,1}}}{\Delta} +
\frac{12+n}{24} 
\frac{E_2 E_4 \widehat{ E_{6,1}}}{\Delta} =
\sum_{N \in {\bf Z} \, or \, {\bf Z}+\frac{3}{4}} d_n
(4N) q^{N} \;\;. \nonumber\\
\eeqa
The world-sheet integral (\ref{intws}) can be evaluated 
using the
techniques of \cite{DKL,HM,K1,K2,NEU}.  A more detailed 
discussion can be found in appendix \ref{wsint}.  Then we find 
from (\ref{fgravv})
that
\beqa
F_1 &=& 24 S - \frac{2}{\pi} \sum_{(k,l,b)>0} 
d_n(4kl-b^2) Li_1 \nonumber\\
&=& 24 {\tilde S} - \frac{1}{\pi} \Big[
\frac{19}{5} \log {\cal C}^2_{30}
+ \frac{3}{5} (1 - 2 n)  \log {\cal C}_5^2 \Big] \;\;.
\label{cons}
\eeqa
Here, we have again ignored the issue of ambiguities linear in $T$, $U$ and
$V$.
Equation (\ref{cons}) gives a 
highly non-trivial consistency check on (\ref{prepstuv}) and on (\ref{intws}).
Namely, it yields, using the product expansions for ${\cal C}_5$
and ${\cal C}_{30}$ given in \cite{GN2} (cf. appendix \ref{prex}),
\beqa
d_n(N)= -\frac{6}{5} N  c_n(N) - \frac{19}{5}
f_2'(N) - \frac{3}{5}(1-2n)f(N) \;\;,
\label{cff}
\eeqa
where $N = 4 kl - b^2 \in 4 {\bf Z} \, or \, 4{\bf Z} + 3$.
As a matter of fact, (\ref{cff}) is equivalent
to the following set of non-trivial relations
\beqa
d^{(1)}_n(N) &=& -\frac{6}{5} N c^{(1)}_n(N) 
- \frac{19}{5}
f_2'(N) - \frac{3}{5}f(N) \;\;, \nonumber\\
d^{(2)}_n(N) &=& -\frac{6}{5} N c^{(2)}_n(N) 
- \frac{1}{5}f(N) \;\;,
\label{c1c2}
\eeqa
where we have decomposed $A_n (4 \tau)$ and $B_n (4 \tau)$ into
\beqa 
A_n (4 \tau)&=& \sum_{N \in 4{\bf Z} \, or \,
4{\bf Z}+3} c^{(1)}_n (N) q^{N} - 6 n \sum_{N \in 4{\bf Z} \, or \, 
4{\bf Z}+3}
c^{(2)}_n (N)_n q^{N} \;\;, \nonumber\\
B_n (4 \tau) &=& \sum_{N \in 4{\bf Z} \, or \, 4{\bf Z}+3} 
d_n^{(1)} (N) q^{N} - 6 n 
\sum_{N \in 4{\bf Z} \, or \, 4{\bf Z}+3}
d_n^{(2)} (N) q^{N} \;\;.
\eeqa
In order to show that (\ref{c1c2}) really holds, consider
introducing \cite{K2}
\beqa
{\hat Z}&=&\frac{1}{72}\frac{(E_4^2 \widehat{E_{4,1}}-
E_6\widehat{E_{6,1}})}{\Delta} \;\;,
\nonumber\\
J_C &=& \frac{2 E_6 \widehat{ E_{6,1}}}{\Delta} + 81 {\hat Z} \;\;,
\eeqa
as well as 
\beqa
{\tilde Z}(\tau) &=& {\hat Z}(4 \tau) = 2 \sum_{N \in 4 {\bf Z} \, or \,
4{\bf Z} + 3} f(N) q^N \;\;, \\ 
{\tilde J_C} (\tau) &=& J_C(4 \tau) = \sum_{N \in 4{\bf Z}, 4{\bf Z}
+ 3} c_J (N) q^{N} = 2 q^{-4} - 14 q^{-1} + 65664 q^3 + 262440 q^4 + \cdots
\;\;. \nonumber
\eeqa
Then, it can be verified that
\beqa
f_2'(N) =  \frac{1}{2} c_J (N) + 6 f(N) \;\;.
\label{f2cjf}
\eeqa
One also has \cite{K2}
\beqa
\Theta_q E_m &=& \frac{m}{12} \left(E_2 E_m - E_{m+2} \right) \;\;,\;\; m=4,6
\nonumber\\
\Theta_q {\hat E}_{m,1} &=& 
\frac{2m-1}{24} \left(E_2 {\hat E}_{m,1} - 
{\hat E}_{m+2,1} \right) \;\;,\;\; m=4,6 \nonumber\\
\Theta_q {\tilde E}_{m,1} &=& 
\frac{2m-1}{6} \left({\tilde E}_2 {\tilde E}_{m,1} - 
{\tilde E}_{m+2,1} \right) \;\;,\;\; m=4,6 
\label{tq}
\eeqa
where 
\beqa
{\tilde E}_2(\tau) &=& E_2 (4 \tau) \;\;, \nonumber\\
{\tilde E}_{m,1}(\tau) &=& {\hat E}_{m,1} (4 \tau) \;\;,
\eeqa
and
where $\Theta_q=q \frac{d}{dq}$.  Then, using (\ref{f2cjf}) as
well as (\ref{tq}), 
it can be shown that
(\ref{c1c2}) indeed holds.

\section{Conclusions}

In this paper we have computed the perturbative threshold corrections,
i.e. the one-loop prepotential and the one-loop gravitational coupling $F_1$,
for $D=4$, $N=2$ heterotic string models compactified on $K3\times T_2$
as a function of the toroidal moduli $T$, $U$ and the single
Wilson line $V$. The considered chain  
of models with generic Abelian gauge group
$U(1)^5$ is characterized by the  embedding of the $SU(2)$
instanton  numbers $(d_1,d_2)=(12-n,12+n)$
into $E_8^{(1)}\times E_8^{(2)}$. At special points in the classical
moduli space ${SO(3,2)\over SO(3)\times SO(2)}/ \Gamma$, where $\Gamma=
SO(3,2,{\bf Z})$ is the classical $T$-duality group, the Abelian gauge group
$U(1)^5$ can be enhanced. The enhancement loci correspond to the Humbert 
surfaces in the classical moduli space. The one-loop prepotential and
the function $F_1$ can be expressed in terms of a set of very beautiful 
modular functions, namely the Siegel and Jacobi modular forms.
The construction of the supersymmetric index
as a power series in the parameter $q=e^{2\pi i\tau}$
involves a so-called hatting
procedure, which describes the transition of going from Jacobi forms
to ordinary modular functions. 
The physical interpretation of the
hatting procedure is just the
turning on of the Wilson line modulus $V$.
If follows that the one-loop prepotential
is given in terms of the same expansion coefficients as the
supersymmetric index.

For the $S$-$T$-$U$-$V$ class of heterotic string models
the spectrum (the
number of massless hyper multiplets) and the 
perturbative threshold corrections explicitly depend on the
particular instanton embedding, parametrized by the integer $n$.
This situation is in contrast to the three parameter $S$-$T$-$U$
class of models, where the spectrum and the 
perturbative couplings do not depend on $n$. In this case the models with 
$n=0,2$ are even equivalent at the non-perturbative level. 
A priori, four-parameter models with gauge group $U(1)^5$ are
obtained for the cases $n=0,1,2$ only. 
In perturbation theory, $E_8^{(1)}$ can only be broken
to some group $G_0^{(1)}$ for $n>2$. However, we
believe that our results also remain valid if there were a mechanism
to get rid of the gauge group $G_0^{(1)}$ as well as of dim($G_0^{(1)}$)
hypermultiplets (leaving 
$215-n$ massless hyper multiplets). 
In fact, for $n=12$ our results perfectly agree
with the recent results of \cite{K2}.

Besides the heterotic construction and the heterotic perturbative
couplings, we also discussed the corresponding dual type II string
models on Calabi--Yau three-folds $X_n$ with Hodge number $h_{1,1}=4$ and 
Euler number $\chi=24n-420$. For $n=0,1,2$ these Calabi--Yau spaces are known
and can be explicitly constructed. 
For $n=2$
there is a Higgs transition \cite{BKKM} to
the three parameter Calabi--Yau $P_{1,1,2,8,12}(24)$;
the possibility of this Higgs transition reflects itself 
in a consistent truncation $V\rightarrow 0$
of the $S$-$T$-$U$-$V$ vector couplings to the corresponding couplings
in the $S$-$T$-$U$ models. 
If the ``complete" gauge symmetry breaking
to $U(1)^5$ on the heterotic side could be realized for $n>2$, it would
predict the existence of new Calabi--Yau spaces $X_n$. Since
the truncation $V\rightarrow 0$ to the perturbative couplings of
three-parameter model consistently works for all $n$, we conjecture
that all Calabi--Yau spaces $X_n$, if existent,  
allow for a Higgs transition to the
three parameter Calabi--Yau spaces. Specifically for $n$ even, the
relevant three parameter Calabi--Yau space should be based on the elliptic
fibration over the Hirzebruch surface $F_2$ (or $F_0$), whereas for
$n$ odd the three parameter Calabi--Yau should be given by the
elliptic fibration over $F_1$. 
The possibility of having  Calabi--Yau spaces $X_n$ with $n>2$ is in fact
supported by the known existence of the $n=12$ Calabi--Yau $P_{2,2,3,3,10}(20)$
\cite{K2}.

Clearly, it would be very interesting to extend these results
to models with a larger number of Wilson lines. 
Finally, it would be very interesting to see if there is any relation
between the perturbative $N=2$ couplings, considered here, 
and the non-perturbative $N=4$
supersymmetric index of \cite{DVV}, where the Siegel modular forms also
play a prominent role.

\vskip1cm

{\bf Acknowlegement}\\

We would like to thank E. Derrick, T. Mohaupt and S. Theisen for discussions.
We would also like to thank B. Andreas 
for participation at an earlier stage of this work, as well as 
P. Berglund and especially P. Mayr for fruitful discussions on the 
model of \cite{BKKM}.

%%%%%%%%%%%%%%%%%%%%% Appendix A  %%%%%%%%%%%%%%%%%%%%%%%%%%%%%%%%%
\appendix

%\begin {appendix}
%{\Large\bf Appendix}\\

\section{Modular forms \label{modforms}}
\setcounter{equation}{0}

\subsection{On Siegel modular forms \label{hump}}

Here we review some properties of Siegel modular forms.
A more detailed account can be found in \cite{I}.

The classical moduli space of a heterotic $S$-$T$-$U$-$V$ model
is locally given by the Siegel upper half plane
${\cal H}_2=\frac{SO(3,2)}{SO(3)\times SO(2)}$ 
(note the exceptional isomorphism
$SO(5)=B_2=C_2=Sp(4)$, here in a noncompact formulation).  The standard
action of $Sp(4,Z)$ on an element $\tau$ of the Siegel upper 
half plane ${\cal H}_2$ is given by 
\beqa
M \rightarrow M\cdot \tau =(a\tau +b)(c\tau +d)^{-1} \;\;,
\eeqa
where
\beqa
\tau =\left( \begin{array}{cc} 
\tau_1 & \tau_3 \\ \tau_3 & \tau_2 
\end{array} \right)
=\left( \begin{array}{cc}  
iT & iV \\ iV & iU 
\end{array} \right) \;\;\;,\;\;\;
M= \left( \begin{array}{cc} a & b \\ c & d \end{array} \right) \;
 \in \;G=Sp(4,Z) \;\;,
\eeqa
and where $\mbox{det}\;\mbox{Im}\tau=\mbox{Re}T\mbox{Re}U-(\mbox{Re}V)^2>0$.
Note that $a,b,c$ and $d$ denote $2 \times 2$ matrices.
A Siegel modular form $F$ of even weight $k$ transforms as
\begin{eqnarray}
F(M\cdot \tau )=det(c\tau +d)^k  F(\tau)
\end{eqnarray}
for every $M \in  \;G=Sp(4,Z)$, whereas a modular form of odd weight $k$
transforms as 
\begin{eqnarray}
F(M\cdot \tau )=\varepsilon (M)det(c\tau +d)^k F(\tau) \;\; .
\end{eqnarray}
Here $\varepsilon :G\rightarrow G/G(2)=S_6\rightarrow \{\pm 1\}$ is the sign
of the permutation in $S_6$.  $G(2)$ denotes the principal
congruence subgroup of level 2.  

The Eisenstein series are given by
\begin{eqnarray}
{\cal E}_k=\sum det (c\tau +d)^{-k} \;\; .
\end{eqnarray}

Now, recall that the usual 
modular forms of $Sl(2,{\bf Z})$ are generated by the 
(normalized) Eisenstein series $E_4$ and $E_6$.  These are 
related to the two
modular forms $E_{12}$ and $\Delta$
of weight $12$ by
\begin{eqnarray}   
aE_4^3+bE_6^2 &=& (a+b)E_{12}  \;\;,  \nonumber\\
E_4^3-E_6^2 &=& \alpha \Delta \;\; ,
\end{eqnarray}
where $\Delta=\eta^{24}$ is the cusp form, and where 
$a=(3\cdot 7)^2, b=2\cdot 5^3, c=a+b=691,
\alpha =2^6\cdot 3^3=1728$.

Similarly,
the ring of Siegel modular forms is generated by the (algebraic independent) 
Eisenstein series
${\cal E}_4, {\cal E}_6, {\cal E}_{10}, {\cal E}_{12}$ and by 
one further
cusp form of odd weight ${\cal C}_{35}$, whose square can again be expressed
in terms of the even generators.
Alternatively, instead of using
${\cal E}_{10}$ and ${\cal E}_{12}$, one can also use  
the cusp forms ${\cal C}_{10}$ and ${\cal C}_{12}$.

A Siegel cusp form is defined as follows.
Since a modular
form $f$ 
is invariant under the translation group $U=\{ \left( \begin{array}{cc}
 1 & b \\ 0 & 1 \end{array} \right) \in G\}$, where the integer valued
$2\times 2$- matrix $b$ is symmetric, it has a Fourier expansion 
$F=\sum_M a(M)e^{2\pi i trM\tau}$.  Here, the summation extends over all 
symmetric half-integral $2\times 2$-matrices (that is,
over symmetric matrices which have 
integer valued diagonal entries and half-integer valued off-diagonal
entries). The
Fourier coefficient $a(M)$ depends only on the class of $M$ under
conjugation by $Sl(2,{\bf Z})$, 
and it is zero unless $M$ is positive semidefinite. 

Now, consider the Siegel operator $\Phi$ which, to every Siegel 
modular
form $F$ 
with Fourier coefficients $a(M)$, associates the ordinary $SL(2,{\bf Z})$ 
modular form $\Phi F$ with Fourier
coefficients 
$a(n)=a(\left( \begin{array}{cc} n & 0 \\ 0 & 0 \end{array} \right))$.
This yields a surjective homomorphism of graded rings of 
modular forms. The forms 
in the kernel are the cusp forms. Thus,
identities between ordinary modular forms
lead to Siegel cusp forms, as follows:
\begin{eqnarray}
E_4E_6=E_{10} &\rightarrow& {\cal E}_4{\cal E}_6-{\cal E}_{10}=:
p \, {\cal C}_{10}     \;\;,   \nonumber\\
aE_4^3+bE_6^2
=cE_{12}&\rightarrow& a{\cal E}_4^3+b{\cal E}_6^2-c{\cal E}_{12}=:
\alpha ^2\frac{ab}{c}{\cal C}_{12}  \;\;, 
\end{eqnarray}
where $p$ denotes a normalisation constant given by 
$p=\frac{2^{10}\cdot 3^5
\cdot 5^2\cdot 7\cdot 53}{43867}$.  We will
drop this normalisation constant in the following, for notational
simplicity.

Next, consider restricting the Siegel modular forms to the 
diagonal $D=\{ \left( \begin{array}{cc} \tau_1 & 0 
\\ 0 & \tau_2 \end{array} \right) \}$ (corresponding to the embedding
$\frac{SO(2,2)}{SO(2)\times SO(2)} 
\rightarrow \frac{SO(3,2)}{SO(3)\times SO(2)}$). Then, interestingly,
\begin{eqnarray}
{\cal E}_k 
\left( \begin{array}{cc} 
\tau_1 & 0 \\ 0 & \tau_2 \end{array} \right)=E_k(\tau_1)E_k(\tau_2) \;\;.
\end{eqnarray}
Specifically
\begin{eqnarray}
{\cal E}_4 &\rightarrow& E_4(\tau_1)E_4(\tau_2) \;\;, \nonumber\\
{\cal E}_6 &\rightarrow& E_6(\tau_1)E_6(\tau_2) \;\;,  \nonumber\\
{\cal C}_{10} &\rightarrow& 0 \;\;, \nonumber\\
{\cal C}_{12} &\rightarrow& \Delta(\tau_1)\Delta(\tau_2) \;\;.
\end{eqnarray}
More precisely, one finds that, up to a normalisation
constant, ${\cal C}_{10} \rightarrow \tau_3^2 \Delta(\tau_1)\Delta(\tau_2)$
as $\tau_3 \rightarrow 0$.

Now,  consider the 
behaviour on $D$ of the 
odd generator ${\cal C}_{35}$.  Since ${\cal C}_{35}$ 
is a more complicated object,  one first reexpresses 
its square in terms of 
the other, even generators.  Namely, by using the results in \cite{I},
one finds that
%------------------------------------------------------------
\begin{eqnarray}
\alpha ^2{\cal C}_{35}^2=\frac{1}{3^3}\C _{10}
&[&\;2^{24}\cdot 3^{15}\C _{12}^5\nonumber\\
& &-2^{13}\cdot 3^9\C _{12}^4(\E _4^3+\E_6^2)\nonumber\\
& &+\;\;\;\;\;\;\;3^3\C _{12}^3
(\E _4^6-2\E _4^3\E _6^2-2^{14}\cdot 3^5\E _4^2\E _6\C _{10}\nonumber\\
& &\;\;\;\;\;\;\;\;\;\;\;\;\;\;\;\;\;\;\;\;-2^{23}3^9 
5^2\E _4\C _{10}^2+\E _6^4)\nonumber\\
& &+2^{11}\cdot3^6\C _{12}^2\C _{10}(37\E _4^4+
5\cdot 7\E _4\E _6^2-2^{12}3^3 5^3\E _6 \C _{10})\nonumber\\
& &+\;\;\;\;\;\;\;3^2\C _{12}
\C _{10}^2(-\E _4^7+2\E _4^4\E _6^2
+2^{11}3^3\cdot 5\cdot 19\E _4^3\E _6\C_{10}\nonumber\\
& &\;\;\;\;\;\;\;\;\;\;\;\;\;\;\;\;\;\;\;\;\;\;
+2^{20}3^6 5^3\cdot 11\E _4^2\C _{10}^2-\E _4\E _6^4
+2^4 3^3 5^2\E _6^3\C _{10})\nonumber\\
& &+\;\;\;\;\;\;\;\;2\cdot\C _{10}^3(-\E _4^4\E _6
-2^{11}3^4\E _4^5\C _{10}+2\E _4^3\E _6^3\nonumber\\
& &\;\;\;\;\;\;\;\;\;\;\;\;\;\;\;\;\;\;\;\;+2^{11}3^4 5^2
\E _4^2\E _6^2\C _{10}+2^{20} 3^7 5^4\E _4\E _6\C _{10}^2-\E _6^5\nonumber\\
& &\;\;\;\;\;\;\;\;\;\;\;\;\;\;\;\;\;\;\;\;\;+2^{31} 3^9 5^5\C _{10}^3)] \;\;.
\end{eqnarray}
%------------------------------------------------------------
Thus, on the diagonal $D$, ${\cal C}_{35} =0$ as well as 
%------------------------------------------------------------
\begin{eqnarray}
\alpha ^2\frac{{\cal C}_{35}^2}{\C _{10}}&=&\alpha ^2[\alpha ^2 \C _{12}^5-
2\C _{12}^4(\E _4^3
+\E _6^2)+\frac{1}{\alpha ^2}\C _{12}^3(\E _4^6-2\E _4^3\E _6^2+\E _6^4)]
\nonumber\\
&=&\C _{12}^3[\alpha ^4\C _{12}^2
-2\alpha ^2\C _{12}(\E _4^3+\E _6^2)+(\E _4^3-\E _6^2)^2]\nonumber\\
&=&\C _{12}^5\frac{(\alpha ^2\C _{12}
-(\E _4^3-\E _6^2))^2-4\alpha^2\C _{12}\E _6^2}{\C _{12}^2} \nonumber\\
&=& {\cal C}_{12}^5 (j(\tau_1) - j(\tau_2))^2 
=(\eta ^2(\tau _1)\eta ^2(\tau _2))^{60}(j(\tau_1)-j(\tau_2))^2 \;\;,
\label{c35c10}
\end{eqnarray}
%------------------------------------------------------------
where
$j(\tau)=E^3_4/\Delta$.  Then, using ${\cal C}_5$ and ${\cal C}_{30}$,
which are related to the forms already defined
by ${\cal C}_{10} = {\cal C}_5^2$ and ${\cal C}_{35} = 
{\cal C}_{30}{\cal C}_{5}$, respectively, it follows that
\beqa
\alpha^2 {\cal C}_{30}^2 \rightarrow \Delta^5(\tau_1) 
\Delta^5(\tau_2) (j(\tau_1)-j(\tau_2))^2  
\label{c30t}
\eeqa
on the diagonal $D$.

A rational quadratic divisor of ${\cal H}_2$ is, by definition \cite{GN2},
the set 
\beqa
{\sc H}_l=\{\left( \begin{array}{cc}  iT & i V \\ i V & i U \end{array} \right)
\in {\cal H}_2|i n_1T+ i m_1U+ i bV+n_2(- TU + V^2)+m_2=0\} \;\;,
\eeqa
where $l=(n_1,m_1,b,n_2,m_2) \in {\bf Z}^5$ is a primitive (i.e.
with the greatest commom divisor equals $1$) integral vector.
The number 
${\rm D}(l)=b^2-4m_1n_1+4n_2m_2$ is called the discriminant of ${\sc H}_l$.
This divisor determines the Humbert surface $H_{\rm D}$ 
in the Siegel three-fold
$Sp_4({\bf Z})\setminus {\cal H}_2$.
The Humbert surface $H_{\rm D}$ 
is (the image in $Sp_4({\bf Z})\setminus {\cal H}_2$
of) the union of all ${\sc H}_l$ of
discriminant ${\rm D}(l)$. 
Each Humbert surface ${\sc H}_{\rm D}$ can be
represented by a linear relation in $T$, $U$ and $V$. 
For instance, the divisor of ${\cal C}_{5}$ is the diagonal 
${\sc H}_{\rm 1}=\{Z=\left( \begin{array}{cc} 
iT & 0 \\ 0 &i U \end{array} \right)
\in Sp_4({\bf Z})\setminus {\cal H}_2\}$. 
Similarly, the divisor of the Siegel
modular form ${\cal C}_{30}$ is the surface
${\sc H}_{\rm 4}=\{Z=\left( 
\begin{array}{cc}i T &i V \\i V &i U \end{array} \right)
\in Sp_4({\bf Z})\setminus {\cal H}_2|T=U\}$. \
The divisor of the Siegel modular form ${\cal C}_{35}$, on the other hand,
is the sum (with multiplicity $1$) of the surfaces 
${\sc H}_{\rm 1}$ and ${\sc H}_{\rm 4}$.

\subsection{On Jacobi forms}

A Siegel modular form $F(T,U,V)$ of weight $k$
has a Fourier expansion with respect to its variable $iU$
%($s={\bf e}[U];T,U\in {\cal H},V\in {\bf C}$)
%-------------------------------------------------------
\begin{eqnarray}
F(T,U,V)=\sum_{m=0}^{\infty}\phi_{k,m}(T,V)s^m  \;\;,
\end{eqnarray}
%--------------------------------------------------------
where $s={\bf e}[iU], {\bf e}[x] = {\rm exp}2 \pi i x$.
Each of the $\phi_{k,m}(T,V)$ is a Jacobi form
of weight $k$ and index $m$ \cite{EZ}.  That is, for each 
$\left( \begin{array}{cc} a & b \\ c & d \end{array} \right)\in
Sl(2,{\bf Z})$ and $\lambda,\mu \in {\bf Z}$
%-------------------------------------------------------
\begin{eqnarray}
\phi_{k,m}(\frac{aT-ib}{icT+d},\frac{V}{icT+d}) &=&
(icT+d)^k e^{2\pi im\frac{c(iV)^2}{icT+d}}\phi(T,V) \;\;,\nonumber\\
\phi_{k,m}(T,V+\lambda T+\mu) &=& e^{-2\pi im(\lambda^2 iT+2\lambda iV)}
\phi_{k,m}(T,V)\;\;.
\end{eqnarray}
%--------------------------------------------------------
A Jacobi form $\phi_{k,m}(T,V)$ of index $m$ has 
in turn 
an
expansion
%-------------------------------------------------------
\begin{eqnarray}
\phi(T,V)=\sum_{n\ge 0}\sum_{l\epsilon {\bf Z}}c(n,l)q^nr^l \;\;,
\end{eqnarray}
%--------------------------------------------------------
where $q={\bf e}[iT], r = {\bf e}[iV]$.
Of special relevance are the Jacobi forms $\phi_{k,1}$ of index $1$.
The summation in $l$ extends
in the usual
case, and for the generators introduced above, over $4n-l^2\ge 0$;
for the forms
divided by $\Delta$, $4n-l^2 \geq  -1 \, {\rm or}\,  -4$, depending on whether
the form is a cusp form or not.
Furthermore
%-------------------------------------------------------
\begin{eqnarray}
c(n,l)=c(4n-l^2) \;\;.
\end{eqnarray}
%--------------------------------------------------------

Consider, for instance, the Eisenstein series, which have 
the expansion
%------------------------------------------------------------
\begin{eqnarray}
{\cal E}_k(T,U,V) = E_k(T) - \frac{2k}{B_k}\, E_{k,1}(T,V)\, s + {\cal O}(s^2)
\;\;.
\end{eqnarray}
%------------------------------------------------------------
Here, the $B_k$ denote the Bernoulli numbers.
Thus, for instance, 
%----------------------------------------------------------
\begin{eqnarray}
{\cal E}_4&=&E_4+240E_{4,1}s+\cdots \;\;, \nonumber\\
{\cal E}_6&=&E_6-504E_{6,1}s+\cdots \;\;.
\end{eqnarray}
%----------------------------------------------------------
The Jacobi forms $E_{4,1}(T,V)$ and $E_{6,1}(T,V)$ 
of index $1$ have the expansion (the expansion coefficients are listed
in the first table of appendix A.6)
%------------------------------------------------------------
\begin{eqnarray}
E_{4,1}&=&1+(r^2+56r+126+56r{-1}+r^{-2})q \nonumber\\
&+&(126r^2+576r+756+576r^{-1}+126r^{-2})q^2+\cdots \;\;, \nonumber\\
E_{6,1}&=&1+(r^2-88r-330-88r^{-1}+r^{-2})q \nonumber\\
&+&(-330r^2-4224r-7524-4224r^{-1}-330r^{-2})q^2+\cdots \;\;.
\end{eqnarray}
%------------------------------------------------------------
Note that 
$E_{k,1}\rightarrow E_k$ as $V \rightarrow 0$.

Similarly, the cusp forms ${\cal C}_{10}(T,U,V)$ and ${\cal C}_{12}(T,U,V)$
have the expansion
\beqa
{\cal C}_{10}(T,U,V) &=& \phi_{10,1}(T,V) s + {\cal O}(s^2) \;\;,\nonumber\\
{\cal C}_{12}(T,U,V) &=& 
\Delta(T) + \frac{1}{12} \phi_{12,1} (T,V) s +  {\cal O}(s^2) \;\;,
\label{c10c12}
\eeqa
where
%------------------------------------------------------------
\begin{eqnarray}
\phi_{10,1}&=&\frac{1}{144}(E_6E_{4,1}-E_4E_{6,1})
\rightarrow \;\;\;\;0 \;\;, \nonumber\\
\phi_{12,1}&=&\frac{1}{144}(E_4^2 E_{4,1}-E_6E_{6,1})\rightarrow 12\Delta
\;\;.
\end{eqnarray}
%------------------------------------------------------------
Here, we have indicated the behaviour under the truncation $V \rightarrow 0$.
The Jacobi forms $\phi_{10,1}$ and $\phi_{12,1}$ of index $1$ have the 
following expansion (the expansion coefficients are listed in the
first table in appendix A.6)

%------------------------------------------------------------
\begin{eqnarray}
\phi_{10,1}&=&(r-2+r^{-1})q+(-2r^2-16r+36-16r^{-1}-2r^{-2})q^2+\cdots \;\;,
\nonumber\\
\phi_{12,1}&=&(r+10+r^{-1})q+(10r^2-88r-132-88r^{-1}+10r^{-2})q^2+\cdots
\;\;.
\end{eqnarray}
%------------------------------------------------------------

\subsection{Product expansions \label{prex}}

The Siegel modular forms ${\cal C}_5$ and 
${\cal C}_{30}={\cal C}_{35}/{\cal C}_{5}$ have the following
product expansion \cite{GN2}
%------------------------------------------------------------
\begin{eqnarray}
{\cal C}_{5}=(qrs)^{1/2}\prod_{n,m,l\in {\bf Z} \atop (n,m,l)>0}
(1-q^nr^ls^m)^{f(4nm-l^2)} \;\;, \nonumber\\
{\cal C}_{30}=(q^3rs^3)^{1/2}(q-s)\prod_{n,m,l\in {\bf Z} \atop (n,m,l)>0}
(1-q^nr^ls^m)^{f'_2(4nm-l^2)} \;\;,
\label{prodexp}
\end{eqnarray}
%------------------------------------------------------------
where the condition $(n,m,l)>0$ means that $n\geq 0,m\geq 0$
and either $l \in {\bf  Z}$
if $n+m>0$, or $l<0$ if $n=m=0$.
The coefficients $f(4nm-l^2)$ and $f'_2(4nm-l^2)$, which are
listed in the first table in appendix A.6,
are defined as follows \cite{GN2}.  Consider the
expansion of
\beqa
\phi_{0,1}:=\frac{\phi_{12,1}}{\Delta (T)}
=\sum_{n\ge 0}\sum_{l\epsilon {\bf Z}}
f(n,l)q^nr^l  \;\;,
\eeqa
where the sum over $l$ is restricted to $4n-l^2\geq -1$.  Then,
$f(N)=f(n,l)$ if $N=4n-l^2\geq-1$, and $f(N)=0$ otherwise.
The coefficients $f_2'(N)$ are then given by
$f'_2(N)=8f(4N)+(2\left (\frac{-N}{2}\right )-3)f(N)+f(\frac{N}{4})$.
Here, $(\frac{D}{2})=1,-1,0$ depending on whether $D\equiv 1 \,
{\rm mod} \, 8, 5\, {\rm mod}\, 8,0\,{\rm mod}\, 2$.

Using the product expansions (\ref{prodexp}), we can perform a check
on the expansion (\ref{c10c12}) of ${\cal C}_{10}=
qrs\prod(1-q^nr^ls^m)^{2f}$.
Namely, consider the term in ${\cal C}_{10}$ with $n=m=0,l=-1$.
It gives rise to 
$qs r(1-r^{-1})^2=qs(r-2+r^{-1})$, which indeed
matches the $q$-term of $\phi_{10,1}$.

Similarly, we can perform a check on (\ref{c30t}).  Setting
$r=1$ in (\ref{prodexp}), we see that the $m=0$-terms have
$f'_2(0)=60$, and thus they match $\Delta ^{5/2}(T) = \eta^{60}$
occuring in 
${\cal C}_{30} \propto \Delta ^{5/2}(T)\Delta ^{5/2}(U)(j(T)-j(U))$.
The sum over $l$ for the terms with $m=n=1$, on the other hand,
yields
$f'_2(4)+2 (f'_2(3)+f'_2(0))=196884$, which matches the $q$-term
in the expansion of $j-744 = q^{-1}+196884q+\cdots$.

\subsection{Theta functions and Jacobi forms \label{thefjaf}}

The 
standard Jacobi theta functions are defined as follows ($z=iV$)
%------------------------------------------------------------
\begin{eqnarray}
\th_1(\tau,z)&=&i\sum_{n\in{\bf Z}}(-1)^nq^{\frac{1}{2}
(n-\frac{1}{2})^2}r^{n-\frac{1}{2}} \;\;,\nonumber\\
\th_2(\tau,z)&=&\sum_{n\in {\bf Z}} 
q^{\frac{1}{2}(n-\frac{1}{2})^2}r^{n-\frac{1}{2}} \;\;,
\nonumber\\
\th_3(\tau,z)&=&\sum_{n\in {\bf Z}} q^{\frac{1}{2}n^2}r^n \;\;,\nonumber\\
\th_4(\tau,z)&=&\sum_{n\in {\bf Z}}(-1)^nq^{\frac{1}{2}n^2}r^n \;\;.
\end{eqnarray}
%------------------------------------------------------------
It is useful to introduce
%------------------------------------------------------------
\begin{eqnarray}
\th_{0,1}(\tau,z)&=&\th_3(2\tau,z)=\sum_{n \in {\bf Z}} q^{n^2}r^n \;\;,
\nonumber\\
\th_{1,1}(\tau,z)&=&\th_2(2\tau,z)=\sum_{n \n {\bf Z}} 
q^{(n-\frac{1}{2})^2}r^{n-\frac{1}{2}}
\end{eqnarray}
%------------------------------------------------------------
as well as
%------------------------------------------------------------
\begin{eqnarray}
\th_{ev}(\tau,z)&=&\th_{0,1}(\tau,2z)=\sum_{n\equiv 0(2)} q^{n^2/4}r^n \;\;,
\nonumber\\
\th_{odd}(\tau,z)&=&\th_{1,1}(\tau,2z)
=\sum_{n\equiv 1(2)} q^{(n-\frac{1}{2})^2}r^{n-\frac{1}{2}} \;\;.
\end{eqnarray}
%------------------------------------------------------------
Next, consider setting $z=0$.  The $\theta_i (\tau,0)$ will
be simply denoted by $\theta_i$, whereas the 
 $\theta_i (2\tau,0)$ will be denoted by $\theta_i(2 \cdot)$
($i=1,\dots 4$).
It is well known that
$\theta_1 = 0$ and that 
$\th_3^4=\th_2^4+\th_4^4$ as well as $\th_2\th_3\th_4=2\eta^3$.
Also
\beqa
E_4&=&\frac{1}{2} \left( \th_2^8 + \th_3^8 + \th_4^8\right) \;\;,\nonumber\\
E_6&=& \frac{1}{2}\left(
\th_2^4+\th_3^4)(\th_3^4+\th_4^4)(\th_4^4-\th_2^4)\right)\nonumber\\
    &=&\frac{1}{2}\left(
-\th_2^6(\th_3^4+\th_4^4)\th_2^2+\th_3^6(\th_4^4-\th_2^4)\th_3^2+
\th_4^6(\th_2^4+\th_3^4)\th_4^2 \right) \;\;.
\eeqa
Additional useful identities are given by
%------------------------------------------------------------
\begin{eqnarray}
2\th_2(2\cdot)\th_3(2\cdot)&=&\th_2^2 \;\;,\nonumber\\
\th_2^2(2\cdot)+\th_3^2(2\cdot)&=&\th_3^2 \;\;,\nonumber\\
\th_3^2(2\cdot)-\th_2^2(2\cdot)&=&\th_4^2 \;\;,\nonumber\\
2\th_2^2(2\cdot)&=&\th_3^2-\th_4^2 \;\;, \nonumber\\
2\th_3^2(2\cdot)&=&\th_3^2+\th_4^2 \;\;,\nonumber\\
\th_4^2(2\cdot)&=&\th_3\th_4 \;\;.
\end{eqnarray}
%------------------------------------------------------------
Now consider 
Jacobi forms 
$f(\tau,z)=\sum_{n\ge 0 \atop l\in {\bf Z}}c(4n-l^2)q^nr^l$
of weight $k$ and index $1$.  The following examples provide 
useful identities 
between Jacobi forms of
index $1$ and Jacobi theta functions
%------------------------------------------------------------
\begin{eqnarray}
\phi_{10,1}&=&\;\;\;\;\;\;\;\;\;\;\;\;\;\;\;\;\;\;\;\;
\;\;\;\;\;\;\;\;\;\;\;\;\;\;\;\;\;\;\;-\eta^{18}\th_1^2(\tau,z) \;\;,
\nonumber\\
\phi_{12,1}&=&12\eta^{24}\frac{\th_3^2(\tau,z)}{\th_3^2}+(\th_4^4-\th_2^4)
[-\eta^{18}\th_1^2(\tau,z)] 
\end{eqnarray}
%------------------------------------------------------------
as well as
%------------------------------------------------------------
\begin{eqnarray}
E_{4,1}&=& \frac{1}{2} \left(
\th_2^6\th_2^2(\tau,z)+\th_3^6\th_3^2(\tau,z)+\th_4^6\th_4^2(\tau,z) 
\right) \;\;, \\
E_{6,1}&=& \frac{1}{2} \left(
-\th_2^6(\th_3^4+\th_4^4)\,\th_2^2(\tau,z)+
\th_3^6(\th_4^4-\th_2^4)\,\th_3^2(\tau,z)+
\th_4^6(\th_2^4+\th_3^4)\,\th_4^2(\tau,z)\right) .  \nonumber
\end{eqnarray}
%------------------------------------------------------------
A Jacobi form of index $1$ 
has the following decomposition \cite{EZ,K1,K2}
%------------------------------------------------------------
\begin{eqnarray}
f(\tau,z)=f_{ev}(\tau)\th_{ev}(\tau,z)+f_{odd}(\tau)\th_{odd}(\tau,z)  \;\;,
\end{eqnarray}
%------------------------------------------------------------
where 
%------------------------------------------------------------
\begin{eqnarray}
f_{ev}&=&\sum_{N\equiv 0(4)}c(N)q^{N/4} \;\;, \nonumber\\
f_{odd}&=&\sum_{N\equiv -1(4)}c(N)q^{N/4} \;\;.
\end{eqnarray}
%------------------------------------------------------------
Consider, for instance, $E_{4,1}$.  It has the decomposition \cite{K2}
\beqa
E_{4,1\,ev }&=& \th_3^7(2\cdot) + 7 \th_3^3(2\cdot)\th_2^4(2\cdot) \;\;,
\nonumber\\  
E_{4,1\,odd }&=& \th_2^7(2\cdot) + 7 \th_2^3(2\cdot)\th_3^4(2\cdot) \;\;.
\eeqa
Furthermore one has (with
 $\theta_{ev} \equiv \theta_{ev}(\tau,z)$
and $\theta_{odd} \equiv \theta_{odd}(\tau,z)$)
%------------------------------------------------------------
\beqa
\th_1^2(\tau,z)&=&\th_2(2\cdot)\th_{ev}-\th_3(2\cdot)\th_{odd}\;\;,\nonumber\\
\th_2^2(\tau,z)&=&\th_2(2\cdot)\th_{ev}+\th_3(2\cdot)\th_{odd}\;\;,\nonumber\\
\th_3^2(\tau,z)&=&\th_3(2\cdot)\th_{ev}+\th_2(2\cdot)\th_{odd}\;\;,\nonumber\\
\th_4^2(\tau,z)&=&\th_3(2\cdot)\th_{ev}-\th_2(2\cdot)\th_{odd} \;\;.
\eeqa
Next, consider the elliptic genus $Z(\tau,z)$ of $K3$,
which is a Jacobi form of weight $0$ and index $1$, given by \cite{KYY}
\beqa
Z(\tau,z)= 2 \frac{\phi_{12,1}}{\Delta}
= 24\frac{\th_3^2(\tau,z)}{\th_3^2}-2\frac{\th_4^4-\th_2^4}{\eta^4}
\frac{\th_1^2(\tau,z)}{\eta^2} \;\;.
\eeqa
It has the decomposition 
\begin{eqnarray}
Z_{ev}&=&24\frac{\th_3(2\cdot)}{\th_3^2}-2\frac{\th_4^4-\th_2^4}{\eta^4}
\frac{\th_2(2\cdot)}{\eta^2} = 20 + 216 q + 1616q^2 + \cdots
\;\;,\nonumber\\
Z_{odd}&=&24\frac{\th_2(2\cdot)}{\th_3^2}+2\frac{\th_4^4-\th_2^4}{\eta^4}
\frac{\th_3(2\cdot)}{\eta^2} = 2 q^{-\frac{1}{4}} - 128 q^{\frac{3}{4}}
- 1026q^{\frac{7}{4}} + \cdots
\;\;.
\end{eqnarray}
Now we introduce the hatted modular function $\widehat{f(\tau,z)}$ as
\beqa
\widehat{f(\tau,z)}=f_{ev}(\tau)+f_{odd}(\tau)\;\;.
\label{fefo}
\eeqa
Hence the hatted modular function corresponds in an one-to-one way
to the index 1 Jacobi form. In particular, the Jacobi form $f(\tau,z)$ and
its hatted relative $\widehat{f(\tau,z)}$ possess identical power
series
expansion coefficients $c(N)$:
\beqa
f(\tau,z)=\sum_{n,l}c(4n-l^2)q^nr^l, \qquad\widehat{f(\tau,z)}=
\sum_{N\in 4{\bf Z} \, or \, 
4{\bf Z} + 3} c(N)q^{N/4} \;\;. \label{hatjacobif}
\eeqa 
Note that an ordinary modular form (that is a form not having any 
$z$-dependence), if 
occuring as a multiplicative
factor in front of a proper Jacobi form, is left
untouched by the hatting procedure (\ref{fefo}).
Thus, for instance,
\beqa
\widehat{E_{4,1}}&=&
\frac{1}{2} \left(\th_2^6\,
\widehat{\th_2^2(\tau,z)}+\th_3^6\,\widehat{\th_3^2(\tau,z)}+
\th_4^6\,\widehat{\th_4^2(\tau,z)}\right) \\
&=&
\frac{1}{2}\left(
\th_2^6[\th_2(2\cdot)+\th_3(2\cdot)]+
\th_3^6[\th_2(2\cdot)+\th_3(2\cdot)]+
\th_4^6[\th_3(2\cdot)-\th_2(2\cdot)]\right) \;\;,\nonumber\\
\widehat{E_{6,1}}&=& \frac{1}{2} \left(
-\th_2^6(\th_3^4+\th_4^4)\,\widehat{\th_2^2(\tau,z)}+
\th_3^6(\th_4^4-\th_2^4)\,\widehat{\th_3^2(\tau,z)}+
\th_4^6(\th_2^4+\th_3^4)\,\widehat{\th_4^2(\tau,z)}\right) \;\;, \nonumber
\eeqa
and similarly
\beqa
{\hat Z} = Z_{ev} + Z_{odd}
=24\frac{\th_2(2\cdot) + \th_3(2\cdot)}{\th_3^2}
-2\frac{(\th_4^4-\th_2^4)}{\eta^4}
\frac{(\th_2(2\cdot)-\th_3(2\cdot))}{\eta^2} 
\;\;.
\eeqa

Furthermore, consider introducing
\beqa
\tilde{f}=\hat{f}(4\cdot)=\sum_{N\in 4{\bf Z} \, or \, 
4{\bf Z} + 3} c(N)q^N \;\;.
\eeqa
Note that $\tilde{f}$ is the $\Gamma_0(4)$
modular form of half-integral weight $k-1/2$ 
associated to a Jacobi form of weight $k$ and index $1$ \cite{EZ}.

\subsection{Lie algebra lattices and Jacobi forms}

The relation between Lie algebra lattice sums 
(see e.g.\cite{LUTHE,LSW})
and Jacobi forms will
be established in three steps. We start by reviewing the well known
relationship between the Lie algebra lattice $E_8$ and the Eisenstein
series $E_4$. Then we go on showing the relation between the 
Lie algebra lattice $E_7$ and the Jacobi Eisenstein series $E_{4,1}$.
Finally, we will relate the processes of splitting off an $A_1$ and the
hatting procedure. This will explain the relation between turning on a
Wilson line and the hatting procedure.

First the relation between the Eisenstein series $E_4$ and the 
partition function of the 
$E_8$ lattice $\Lambda=\{x\in{\bf Z}^8\cup\pi+{\bf Z}^8|(x,\pi)\in{\bf Z}\}$
is well known ($\pi=(1/2,\cdots,1/2)\in{\bf Z}^8$) and reads
%------------------------------------------------------------
\begin{eqnarray}
E_4=\sum_{x\in\Lambda}q^{\frac{1}{2}x^2}=\frac{1}{2}(\th_2^8+\th_3^8+\th_4^8)
\;\;.
\end{eqnarray}
%------------------------------------------------------------
Because of the lattice relation 
$\Lambda_{E_8}=\Lambda_{D_8^{(0)}}+\Lambda_{D_8^{(S)}}$, this also 
shows that
the fermionically computed partition function  
$P_{D_8^{(0)}}+P_{D_8^{(S)}}$ of $E_8$ is identical to 
the bosonically computed one, if one recalls the relation
between the bosonic conjugacy class picture and the 
fermionic boundary condition picture
%------------------------------------------------------------
\begin{eqnarray}
P_{D_n^{(0)}}&=&\frac{\th_3^n+\th_4^n}{2}=\frac{NS^+ +NS^-}{2} \;\;,\nonumber\\
P_{D_n^{(V)}}&=&\frac{\th_3^n-\th_4^n}{2}=\frac{NS^+ -NS^-}{2} \;\;,\nonumber\\
P_{D_n^{(S/C)}}&=&\frac{\th_2^n}{2}=\frac{R^+}{2} \;\;.
\end{eqnarray}
%------------------------------------------------------------
Now consider the Jacobi form $E_{4,1}(\tau,z)=\sum c(4n-l^2)q^nr^l$.
Since the expression $\sum_{x\in\Lambda}q^{\frac{1}{2}x^2}r^{(x,\pi)}$
has the correct weights (and truncation), and since the space in question is 
one--dimensional, this 
represents $E_{4,1}$. If one considers the $l=0$ resp. $l=1$ sector, one
finds $\sum_{(x,\pi)=0}q^{\frac{1}{2}(x,x)}=\sum_n c(4n)q^n=
\sum_{N\equiv 0(4)}c(N)q^{N/4}$ resp. $\sum_{(x,\pi)=1}q^{\frac{1}{2}(x,x)}=
\sum_n c(4n-1)q^n=\sum_{N\equiv -1(4)}c(N)q^{\frac{N+1}{4}}$, i.e.
$E_{4,1ev}=\sum_{(x,\pi)=0}q^{\frac{1}{2}x^2}$ and
$E_{4,1odd}=q^{-1/4}\sum_{(x,\pi)=1}q^{\frac{1}{2}x^2}=
\sum_{x\in -\frac{\pi}{2}+\Lambda \atop (x,\pi)=0}q^{\frac{1}{2}x^2}$.
Thus, 
%------------------------------------------------------------
\begin{eqnarray}
E_{4,1 \,ev}=\sum_{(x,\pi)=0}^{E_8}q^{\frac{1}{2}x^2}&=&
\sum_{x\in (0)}^{E_7}q^{\frac{1}{2}x^2}=P_{E_7^{(0)}} \;\;, \nonumber\\
E_{4,1 \,odd}=\sum_{x\in -\frac{\pi}{2}+\Lambda \atop (x,\pi)=0}^{E_8}
q^{\frac{1}{2}x^2}&=&\sum_{x\in (1)}^{E_7}q^{\frac{1}{2}x^2}=P_{E_7^{(1)}}\;\;,
\end{eqnarray}
%------------------------------------------------------------
where the lattice sums 
$P_{E_7^{(i)}}=\sum_{x\in (i)}q^{\frac{1}{2}x^2}$ 
run over vectors within the conjugacy class (i).\\
Besides this lattice theoretic argument, this can also be checked explicitely
%------------------------------------------------------------
\begin{eqnarray}
E_{4,1 \,ev}&=&\th_3^3(2\cdot)
(\th_3^4(2\cdot)+7\th_2^4(2\cdot))=\th _3(2\cdot)\th _3^2(2\cdot)
(\th _4^4(2\cdot)+8\th _2^4(2\cdot))\nonumber\\
&=&\th _3(2\cdot)\frac{\th_3^2+\th_4^2}{2}[\th_3^2\th_4^2+2(\th_3^2-\th_4^2)^2]
=\th _3(2\cdot)[\th_3^6+\th_4^6-\frac{\th_3^2\th_4^2}{2}
(\th_3^2+\th_4^2)]\nonumber\\
&=&\th _3(2\cdot)\frac{1}{2}[\th_3^6+\th_4^6]+\th_2(2\cdot)\frac{1}{2}\th_2^6
=P_{E_7^{(0)}} \;\; ;\label{efoureven}
\end{eqnarray}
%------------------------------------------------------------
 similarly $E_{4,1 \,odd}= P_{E_7^{(1)}}$.  
 
The last relation in (\ref{efoureven}) follows
by noting the following lattice decomposition of $P_{E_7^{(0)}}$:
%$\Lambda_{E_7^{(0)}}=\Lambda_{D_6^{(0)}}\oplus\Lambda_{A_1^{(0)}}+
%(\Lambda_{D_6^{(S)}};\frac{1}{2}\Lambda_{A_1^{(0)}})$ implies 
$P_{E_7^{(0)}}=P_{D_6^{(0)}}\cdot P_{A_1^{(0)}}+
P_{D_6^{(S)}}\cdot P_{A_1^{(1)}}$.
Here one uses the following lattice sums for $A_1$,
which 
has the root lattice
$\Lambda_{A_1}^{(0)}=\sqrt{2}{\bf Z}$ and two conjugacy classes:  
%------------------------------------------------------------
\begin{eqnarray}
P_{A_1^{(0)}}&=& \sum_{x\in (0)}^{A_1}
q^{\frac{1}{2}x^2}=\sum_{n\in {\bf Z}} q^{n^2}=
\th_3(2\cdot)\;\;,\nonumber\\
P_{A_1^{(1)}}&=&\sum_{x\in (1)}^{A_1}q^{\frac{1}{2}x^2}=
\sum_{n\in {\bf Z}}  q^{(n-1/2)^2}=
\th_2(2\cdot) \;\;.
\end{eqnarray}
%------------------------------------------------------------

Thus we get that
\begin{eqnarray}
2\widehat{E_{4,1}}&=&\th_2^6[\th_2(2\cdot)+\th_3(2\cdot)]+
\th_3^6[\th_2(2\cdot)+\th_3(2\cdot)]+
\th_4^6[\th_3(2\cdot)-\th_2(2\cdot)]\nonumber\\
&=&\th_2^6\cdot\widehat{\th_2^2(\tau,z)}+\th_3^6\cdot\widehat{\th_3^2(\tau,z)}+
\th_4^6\cdot\widehat{\th_4^2(\tau,z)}\nonumber\\
&=&2(P_{E_7^{(0)}}+P_{E_7^{(1)}}) 
\;\; ,
\end{eqnarray}
%------------------------------------------------------------
which also holds, as is easily seen, in the dehatted version.
Now we understand that the breaking of $E_8$ to $E_7$ by turning
on a Wilson line, i.e.
the splitting off of an $A_1^{\rm{Wilson}}$, precisely corresponds
to the replacement of $E_4$ by the hatted modular function $\widehat{E_{4,1}}$.

On the other hand, note that the truncation $V\rightarrow 0$ 
%------------------------------------------------------------
\begin{eqnarray}
E_{4,1}(\tau,0)=E_4=(E_{4,1})_{ev}\th_3(2\cdot)+(E_{4,1})_{odd}\th_2(2\cdot)
\end{eqnarray}
%------------------------------------------------------------
reflects the decomposition of $E_8 \supset E_7 \times A_1$
%------------------------------------------------------------
\begin{eqnarray}
P_{E_8}=P_{E_7^{(0)}}\cdot P_{A_1^{(0)}}+P_{E_7^{(1)}}\cdot P_{A_1^{(1)}}
\;\;.
\end{eqnarray}
%------------------------------------------------------------
%Similarly, (A.47)
%reflects a decomposition related to $E_7$ under $D_6 \times A_1$,
%when rewritten as
%------------------------------------------------------------
%\begin{eqnarray}
%\widehat{E_{4,1}} =P_{E_7^{(0)}}+P_{E_7^{(1)}}=
%P_{D_6^{(0)}}\cdot P_{A_1^{(0)}}+P_{D_6^{(V)}}\cdot P_{A_1^{(1)}}+
%P_{D_6^{(S/C)}}\cdot P_{A_1^{(0)}}+P_{D_6^{(C/S)}}\cdot P_{A_1^{(1)}}
%\;\;. \nonumber\\
%\end{eqnarray}
%------------------------------------------------------------

Let us again demonstate the hatting procedure by considering
the Wilson line breaking of 
$D_2=A_1 \times 
A_1^{\rm{Wilson}}$ to $A_1$.
The lattice decomposition of $D_2$ under $A_1\times A_1$ has the form
\begin{eqnarray}
P_{D_2^{(0)}}&=&{\th_3^2+\th_4^2\over 2}=P_{A_1^{(0)}}\cdot P_{A_1^{(0)}}=
\th_3(2\cdot )^2,\nonumber\\
P_{D_2^{(V)}}&=&{\th_3^2-\th_4^2\over 2}=P_{A_1^{(1)}}\cdot P_{A_1^{(1)}}=
\th_2(2\cdot )^2,\nonumber\\
P_{D_2^{(S,C)}}&=&{\th_2^2\over 2}=P_{A_1^{(0)}}\cdot P_{A_1^{(1)}}=
\th_2(2\cdot )\th_3(2\cdot ).
\end{eqnarray}
Thus the corresponding hatted Jacobi forms become
\begin{eqnarray}
{\widehat{\th_3^2(\tau,z)}+\widehat{\th_4^2(\tau,z)}
\over 2}&=&P_{A_1^{(0)}}=
\th_3(2\cdot ),\nonumber\\
{\widehat{\th_3^2(\tau,z)}-\widehat{\th_4^2(\tau, z)}
\over 2}&=&P_{A_1^{(1)}}=
\th_2(2\cdot ),\nonumber\\
{\widehat{\th_2^2(\tau,z)}\over 2}&=&{1\over 2}(P_{A_1^{(0)}}+ P_{A_1^{(1)}})=
{1\over 2}(\th_2(2\cdot )+\th_3(2\cdot )).
\end{eqnarray}

Finally, 
going back from the conjugacy class picture to the boundary condition
picture   one has 
\begin{eqnarray}
NS^{\pm}_{A_1}=P_{A_1^{(0)}}\pm P_{A_1^{(1)}}=\th_3(2\cdot)\pm \th_2(2\cdot)
=\widehat{\th_{3/4}^2(\tau,z)} \;\;, \\
R^+_{A_1}=P_{A_1^{(0)}}+ P_{A_1^{(1)}}=\th_3(2\cdot)+ \th_2(2\cdot)
=\widehat{\th_{2}^2(\tau,z)} \;\; .
\end{eqnarray}

\subsection{Tables}

This table displays some expansion coefficients of the Jacobi forms
$E_{4,1}$, $E_{6,1}$, $\phi_{10,1}$, $\phi_{12,1}$ and of the Siegel
forms ${\cal C}_5$, ${\cal C}_{30}$.

\begin{tabular}{|c||c|c|c|c||c|c|} \hline
N &$e_{4,1}(N)$&$e_{6,1}(N)$&$c_{10,1}(N)$&$c_{12,1}(N)$&$f(N)
$&$f_2^{\prime}(N)$\\ \hline \hline
-4&      - &      - &         - &         - & -    &   1  \\
-1&      - &      - &         - &         - &  1   &  -1  \\
0 &      1 &      1 &         0 &         0 & 10   &  60  \\
3 &     56 &    -88 &         1 &         1 &-64   &32448 \\
4 &    126 &   -330 &        -2 &        10 &108   &131868\\
7 &    576 &  -4224 &       -16 &       -88 &-513  & ***  \\
8 &    756 &  -7524 &        36 &      -132 & 808  & ***  \\
11&   1512 & -30600 &        99 &      1275 &-2752 & ***  \\
12&   2072 & -46552 &      -272 &       736 & 4016 & ***  \\
15&   4032 &-130944 &      -240 &     -8040 &-11775& ***  \\
16&   4158 &-169290 &      1056 &     -2880 & 16524& ***  \\
19&   5544 &-355080 &      -253 &     24035 &  *** & ***  \\
20&   7560 &-464904 &     -1800 &     13080 &  *** & ***  \\ \hline
\end{tabular}

\hspace{0.5cm}

In the following table some expansion coefficients of ${E_{4,1}E_6\over \Delta}$,
${E_4E_{6,1}\over \Delta}$ and of $A_n$ (see eq.(\ref{An})) for $n=0,1,2,12$
are listed.

\hspace{0.5cm}

\begin{tabular}{|c||c|c||c|c|c|c|} \hline
N & $E_{4,1}E_6/\Delta$ & $E_4E_{6,1}/\Delta$ 
&$2A_0$&$2A_1$&$2A_2$&$2A_{12}$\\ \hline \hline
-4 &  1  &  1 & 2 & 2 & 2 & 2\\
-1 & 56  & -88  & -32 & -44 & -56 & -176\\
 0 &-354 & -66  & -420 & -396 & -372 & -132\\
 3 &-26304 & -27456  & -52760 & -53356 & -53952 & -54912\\
 4 &-88128 & -86400  & -174528 & -174384 & -174240 & -172800\\ \hline
\end{tabular}

%%%%%%%%%%%%%%%%%%% Appendix B %%%%%%%%%%%%%%%%%%%%%%%%%%%

\section{The world sheet integral ${\tilde I}_{3,2}$ \label{wsint}}
\setcounter{equation}{0}

Consider the integral
\beqa
{\tilde I}_{3,2} &=& 
\int_{\cal F} 
\frac{d^2\tau}{\tau_2} \Big[ Z_{3,2} F(\tau) \Big(E_2 - \frac{3}{\pi
\tau_2} \Big)
 -d_n(0) \Big] \;\;,
\label{tildei32}
\eeqa
where 
\beqa
F(\tau) &=& A_n = \sum_{N \in {\bf Z} , {\bf Z} + \frac{3}{4}}
c_n(4N) q^N \;\;, \nonumber\\
B_n(\tau) &=& A_n E_2 = \sum_{N \in {\bf Z} , {\bf Z} + \frac{3}{4}}
d_n(4N) q^N \;\;.
\eeqa
${\cal F}$ denotes the fundamental domain for $SL(2,{\bf Z})$.

The calculation of (\ref{tildei32}) involves three
contributions \cite{DKL,HM,K1,K2,NEU}, that is
${\tilde I}_{3,2} = {\cal I}_0 + {\cal I}_{nd} + {\cal I}_{deg}$.
In this appendix, we will evalute ${\cal I}_{nd}$ by closely
following the procedure described in \cite{DKL,HM,K1,K2,NEU}.
We will work in the chamber $T_2 > U_2 > 2V_2$.  The other two contributions
can be evaluated along similar lines.

Recall that 
\beqa
Z_{3,2}(\tau,{\bar \tau}) = \sum_{p \in \Gamma^{3,2}}  q^{\frac{1}{2}p^2_L}
 {\bar q}^{\frac{1}{2}p^2_R} = \sum_{m_1,m_2,n_1,n_2,b}
  q^{\frac{1}{2}(p^2_L-p^2_R)}
  q^{\frac{1}{2}p^2_R}
 {\bar q}^{\frac{1}{2}p^2_R} \;\;,
\eeqa
where
\beqa
p^2_R &=&
 \frac{|m_2 + m_1 U + n_1 T + n_2 (TU - V^2) +  b V|^2}{2Y} \;\;, \nonumber\\
\frac{1}{2}(p^2_L - p^2_R) &=& \frac{1}{4} b^2-m_1n_1+m_2n_2  \;\;, \nonumber\\
Y &=& T_2 U_2 - V^2_2  > 0 \;\;.
\eeqa

Performing a Poisson resummation on $m_1$ and $m_2$ yields \cite{HM,NEU}
\beqa
\sum_{m_1,m_2}  q^{\frac{1}{2}p^2_R}
 {\bar q}^{\frac{1}{2}p^2_R} = \sum_{k_1,k_2} \frac{Y}{U_2 \tau_2} 
q^{\frac{b^2}{4}}
e^{\cal G}
\;\;,
\eeqa
where
\beqa
{\cal G} &=& - \frac{\pi Y}{U_2^2 \tau_2} |{\cal A}|^2 - 2 \pi i T \det A
+ \frac{\pi b}{U_2} \left(V {\tilde {\cal A}} - {\bar V} {\cal A} \right)
\nonumber\\
&-& \frac{\pi n_2}{U_2} \left( 
V^2 {\tilde {\cal A}} - {\bar V}^2 {\cal A} \right)
+ \frac{2 \pi i V^2_2}{U_2^2} ( n_1 + n_2 {\bar U}) {\cal A} \;\;.
\eeqa
Here, 
\beqa
A &=&\left( \begin{array}{cc}  n_1 &  -k_1 \\  n_2 &  k_2 \end{array} 
\right) \;\;,   \nonumber\\
{\cal A} &=& (1,U)A
(\tau,1)^T = 
-k_1 + n_1 \tau + k_2 U + n_2 \tau U \;\;, \nonumber\\
{\tilde {\cal A}} &=& (1,{\bar U})A
(\tau,1)^T = 
- k_1 + n_1 \tau + k_2 {\bar U} + n_2 \tau {\bar U} \;\;.
\eeqa
The contribution
${\cal I}_{nd}$ is obtained by restriction to 
 non-denerate matrices $A$ (that is, matrices with non-zero determinant)
of the form \cite{DKL,HM}
\beqa
A=
\left( \begin{array}{cc}  n_1 &  -k_1 \\  0 &  k_2 \end{array} 
\right) \equiv
\left( \begin{array}{cc}  k &  j \\  0 &  p \end{array} 
\right) \;\;\;,\;\;  p \neq 0 \;,\; k > j \geq 0 \;\;.
\eeqa
Then \cite{DKL,HM}
\beqa
{\cal I}_{nd} = 2 \frac{Y}{U_2} \sum_{b \in {\bf Z} }
\sum_{p \in {\bf Z} \atop  p \neq 0}
\sum_{k >0} 
\sum_{j=0}^{k-1}
\int_{- \infty}^{\infty} d \tau_1
\int_{0}^{\infty} \frac{d \tau_2}{\tau^2_2} q^{\frac{b^2}{4}}
e^{\cal G} F(\tau) \Big(E_2 - \frac{3}{\pi
\tau_2} \Big) \;\;,
\label{intnd}
\eeqa
where
\beqa
{\cal G} &=& - 2 \pi i T kp - \frac{\pi Y}{\tau_2 U_2^2} (k\tau_2 + p U_2)^2
- 2 \pi b k \frac{V_2}{U_2} \tau_2
\nonumber\\
&+& 2 \pi i 
\frac{b }{U_2} ( j V_2 -  p V_1 U_2 + pU_1V_2)
\nonumber\\
&+& 2 \pi i \frac{V_2^2}{U_2^2} k (j + i k \tau_2 + p U) \nonumber\\
&-& \frac{\pi Y}{\tau_2 U_2^2}k^2 (\tau_1 + \frac{j+pU_1}{k})^2
+ 2 \pi i \frac{V_2}{U_2} b k \tau_1 + 2 \pi i \frac{V_2^2}{U_2^2} k^2
\tau_1 \;\;.
\eeqa
The integral over $\tau_1$ is gaussian
and yields
\beqa
\int_{- \infty}^{\infty} d \tau_1 
e^{- \frac{\pi Y}{\tau_2 U_2^2}k^2 (\tau_1 + \frac{j+pU_1}{k})^2 + 
 2 \pi i \tau_1 {\tilde N}} = 
\frac{U_2}{k}  \sqrt{\frac{\tau_2}{Y}}
e^{- \frac{\pi \tau_2 U_2^2}{Y k^2} {\tilde N}^2 -
 2 \pi i \frac{j + p U_1}{k} {\tilde N}} \;\;,
\eeqa
where
\beqa
{\tilde N} = N + \frac{b^2}{4} + bk \frac{V_2}{U_2} + k^2 \frac{V_2^2}{U_2^2}
\;\;.
\eeqa
Then, ${\cal I}_{nd}$ turns into 
\beqa
{\cal I}_{nd} &=& 2 \sqrt{Y} \sum_{N \in {\bf Z}, {\bf Z + 
\frac{3}{4}}} \sum_{b \in {\bf Z} }\;
\sum_{p \in {\bf Z} \atop  p \neq 0}\;
\sum_{k >0} \frac{1}{k}\;
\sum_{j=0}^{k-1} \; \nonumber\\
&& \int_{0}^{\infty} \frac{d \tau_2}{\sqrt{\tau^3_2}}
e^{\cal G'}  e^{- 2 \pi \tau_2 (N + \frac{b^2}{4})}
e^{- \frac{\pi \tau_2 U_2^2}{Y k^2} {\tilde N}^2 -
 2 \pi i \frac{j + p U_1}{k} {\tilde N}} 
\Big( d_n(4N)  - \frac{3c_n(4N)}{\pi
\tau_2} \Big) \;\;,
\eeqa
where
\beqa
{\cal G}' &=& - 2 \pi i T kp - \frac{\pi Y}{\tau_2 U_2^2} (k\tau_2 + p U_2)^2
- 2 \pi b k \frac{V_2}{U_2} \tau_2
\nonumber\\
&+& 2 \pi i 
\frac{b }{U_2} ( j V_2 -  p V_1 U_2 + pU_1V_2)
+ 2 \pi i \frac{V_2^2}{U_2^2} k (j + i k \tau_2 + p U) \;\;.
\eeqa
Next, consider summing over $j$. Then
\beqa
\sum_{j=0}^{k-1} 
e^{- 2 \pi i \frac{j}{k} {\tilde N}  + 2 \pi i  
\frac{V_2 }{U_2} jb  + 2 \pi i \frac{V_2^2}{U_2^2} kj} 
=  \sum_{j=0}^{k-1} 
e^{- 2 \pi i \frac{j}{k} (N + \frac{b^2}{4})}
= \Biggl \lbrace  \begin{array}{c}  k \;\;
{\rm if} \; \frac{N + \frac{b^2}{4}}{k} = l \in {\bf Z} \;\;. \\
0 \;\; {\rm otherwise}\;\;. \end{array} 
\eeqa 
Note that setting $N=kl - \frac{b^2}{4}$ is consistent with
$N \in {\bf Z}, {\bf Z} + \frac{3}{4}$.
It follows that 
\beqa
{\cal I}_{nd} &=& 2 \sqrt{Y} \sum_{l \in {\bf Z}} 
\sum_{b \in {\bf Z} }\;
\sum_{p \in {\bf Z} \atop  p \neq 0}\;
\sum_{k >0} \; \\
&& \int_{0}^{\infty} \frac{d \tau_2}{\sqrt{\tau^3_2}}
e^{\cal G''}  e^{- 2 \pi \tau_2 kl}
e^{- \frac{\pi \tau_2 U_2^2}{Y k^2} {\tilde N}^2 -
 2 \pi i \frac{j + p U_1}{k} {\tilde N}} 
\Big( d_n(4 kl -  b^2 )  - \frac{3c_n(4 kl - b^2)}{\pi
\tau_2} \Big) , \nonumber
\eeqa
where now ${\tilde N} = k(l + b \frac{V_2}{U_2} + k \frac{V_2^2}{U^2_2})$,
and where
\beqa
{\cal G''} &=& - 2 \pi i T kp - \frac{\pi T_2}{\tau_2 U_2} 
(k\tau_2 + p U_2)^2
- 2 \pi b k \frac{V_2}{U_2} \tau_2 - \pi k^2 \frac{V^2_2}{U^2_2} \tau_2
+ \frac{\pi V_2^2}{\tau_2} p^2
\nonumber\\
&+& 2 \pi i 
\frac{bp }{U_2} ( -   V_1 U_2 + U_1V_2)
+ 2 \pi i \frac{V_2^2}{U_2^2} U_1 p k  \;\;.
\eeqa
Next, rewrite the sum over $p \neq 0 $ as
\beqa
{\cal I}_{nd} &=& 2 \sqrt{Y} \sum_{l \in {\bf Z}} 
\sum_{b \in {\bf Z} }\;
\sum_{p > 0}\;
\sum_{k >0} \; \nonumber\\
&&\Big( e^{2 \pi i T kp + 2 \pi i U_1 pl + 2 \pi i V_1 pb}
+ e^{- 2 \pi i {\bar T} kp  - 2 \pi i U_1 pl - 2 \pi i V_1 pb} \Big)
e^{2 \pi kp T_2} 
\nonumber\\
&& \int_{0}^{\infty} \frac{d \tau_2}{\sqrt{\tau^3_2}}
  e^{-  A  \tau_2  }
e^{-  \frac{B}{\tau_2}}
\Big( d_n(4 kl -  b^2 )  - \frac{3c_n(4 kl - b^2)}{\pi
\tau_2} \Big) ,
\eeqa
where
\beqa
A &=& \pi (2 kl + 2 b k \frac{V_2}{U_2} + \frac{U_2^2}{Y k^2}
{\tilde N}^2 + k^2 \frac{T_2}{U_2} + k^2 \frac{V^2_2}{U^2_2}
 ) = \frac{\pi}{Y} 
(k T_2 + l U_2 + b V_2 )^2 
 \;\;, \nonumber\\
B&=& \pi p^2 Y \;\;.
\eeqa
Then, by using the following integral representations for the
Bessel functions $K_{\frac{1}{2}}$ and $K_{\frac{3}{2}}$ (for $A >0$, $B>0$)
\beqa
\int_{0}^{\infty} \frac{d \tau_2}{\sqrt{\tau_2^3}}
e^{- A \tau_2} e^{-\frac{B}{\tau_2}} &=& \sqrt{ \frac{\pi}{B}} e^{-2 \sqrt{AB}}
\;\;,\nonumber\\
\int_{0}^{\infty} \frac{d \tau_2}{\sqrt{\tau_2^5}}
e^{- A \tau_2} e^{-\frac{B}{\tau_2}} &=& \frac{\sqrt \pi}{B} 
e^{- 2\sqrt{AB}} \left(\sqrt{A} + \frac{1}{2 \sqrt{B}} \right) \;\;,
\eeqa
it follows that 
\beqa
{\cal I}_{nd} &=& 2  \sum_{l \in {\bf Z}} 
\sum_{b \in {\bf Z} }\;
\sum_{p > 0}\;
\sum_{k >0} \; \Big( e^{2 \pi i p r \odot y}
+
e^{- 2 \pi i p r \odot y}
\Big)
\nonumber\\
&& 
\Bigg[ \frac{d_n(4 kl -  b^2 )}{p}  - \frac{3 c_n(4 kl - b^2)}{\pi Y}
(\frac{|kT_2 + lU_2 + bV_2|}{p^2} + \frac{1}{2 \pi p^3}) 
 \Bigg] ,
\eeqa
where
\beqa
r \odot y = k T_1 + l U_1 + b V_1 + i |k T_2 + l U_2 + b V_2| \;\;.
\eeqa
Note that, in the chamber $T_2 > U_2 > 2 V_2$,
$|kT_2 + lU_2 + bV_2|= kT_2 + lU_2 + bV_2$
and, hence,
$r \odot y = kT + lU + b V$. This is due to the fact that the coefficients
$c_n(4kl-b^2)$ and $d_n(4kl-b^2)$ vanish unless $4kl-b^2\geq-4$.

Then, summing over $p$ yields
\beqa
{\cal I}_{nd} &=& 4 \Re \Bigg(  \sum_{l \in {\bf Z}} 
\sum_{b \in {\bf Z} }\;
\sum_{k >0} \; 
\Bigg[ d_n(4 kl -  b^2 ) Li_1(e^{2 \pi i (kT + lU + bV)})  \nonumber\\
&-& \frac{3}{\pi Y} c_n(4 kl - b^2) {\cal P} (e^{2 \pi i (kT + lU + bV)}) 
 \Bigg] \Bigg),
\eeqa
where we introduced \cite{HM}
\beqa
{\cal P} (e^{2 \pi i (kT + lU + bV)}) = ( kT_2 + lU_2 + bV_2) Li_2(
e^{2 \pi i (kT + lU + bV)}) + \frac{1}{2\pi} Li_3(
 e^{2 \pi i (kT + lU + bV)}) .
\eeqa
The term proportional to $\frac{1}{Y} c_n{\cal P}$ contributes to the
Green--Schwarz term \cite{HM}, whereas the term proportional to $d_n Li_1$ 
contributes to $F_1$.

%%%%%%%%%%%%%%%%%%%%%%%%%%%%%%%%%%%%%%%%%%%%%%%%%%


\begin{thebibliography}{99}

\newcommand{\NP}[3]{{ Nucl. Phys.} {\bf #1} {(19#2)} {#3}}
\newcommand{\PL}[3]{{ Phys. Lett.} {\bf #1} {(19#2)} {#3}}
\newcommand{\PRL}[3]{{ Phys. Rev. Lett.} {\bf #1} {(19#2)} {#3}}
\newcommand{\PR}[3]{{ Phys. Rev.} {\bf #1} {(19#2)} {#3}}
\newcommand{\IJ}[3]{{ Int. J. Mod. Phys.} {\bf #1} {(19#2)}
  {#3}}
\newcommand{\CMP}[3]{{ Comm. Math. Phys.} {\bf #1} {(19#2)} {#3}}
\newcommand{\PRp} [3]{{ Phys. Rep.} {\bf #1} {(19#2)} {#3}}



\bibitem{sduality} A.~Font, L.~Ib\'a\~nez, D.~L\"ust and F.~Quevedo,
  Phys. Lett. {\bf B 249} (1990) 35;\\
  S.--J.~Rey, Phys. Rev. {\bf D 43} (1991) 526;\\
  A.~Sen, Phys. Lett. {\bf B 303} (1993) 22, {\bf
B 329} (1994) 217;\\
  J.~Schwarz and A.~Sen,  Nucl. Phys.
{\bf B 411} (1994) 35, hep-th/9304154.

\bibitem{SchSen}
J. Schwarz and A. Sen, \PL{B 312}{93}{105}, hep-th/9305185.

\bibitem{Sen}
A. Sen, \IJ{A 9}{94}{3707}, hep-th/9402002.

\bibitem{HullTown}
C. M. Hull and P. Townsend, \NP{B 438}{95}{109}, hep-th/9410167.

\bibitem{Wit1}
E. Witten,  \NP{B 443}{95}{85}, hep-th/9503124.

\bibitem{KV} S. Kachru and C. Vafa, Nucl. Phys. {\bf B 450} (1995) 
69, hep-th/9505105.

\bibitem{KKLMV} 
G. L. Cardoso, D. L\"ust and T. Mohaupt, \NP{B 455}{95}{131}, 
hep-th/9507113;\\ 
S. Kachru, A. Klemm, W. Lerche, P. Mayr and
C. Vafa, \NP{B 459}{96}{537}, hep-th/9508155;\\
I. Antoniadis and H. Partouche, \NP{B 460}{96}{470}, hep-th/9509009;  \\
A. Klemm, P. Mayr, W. Lerche, C. Vafa and N. Warner, hep-th/9604034. 

\bibitem{SW} N. Seiberg and E. Witten, \NP{B 426}{94}{19}, hep-th/9407087; 
\NP{B 431}{94}{484}, hep-th/9408099;\\
A. Klemm, W. Lerche, S. Theisen and
S. Yankielowicz, \PL{B 344}{95}{169},
hep-th/9411048;\\
P. Argyres and A. Faraggi, \PRL{74}{95}{3931},  hep-th/9411057; \\
A. Klemm,
W. Lerche and S. Theisen, \IJ{A 11}{96}{1929},  hep-th/9505150.

\bibitem{nonedual}
C. Vafa and E. Witten, hep-th/9507050;\\
J. Harvey, D. Lowe and A. Strominger, Phys. Lett. {\bf B 362} (1995) 65, 
hep-th/9507168;\\ 
E. Witten, hep-th/960430;\\
S. Sethi, C. Vafa and E. Witten, hep-th/9606122;\\
I. Brunner and R. Schimmrigk, hep-th/9606148;\\
R. Gopakumar and S. Mukhi, hep-th/9607057;\\
R. Donagi, A. Grassi and E. Witten, hep-th/9607091;\\
M. Bianchi, S. Ferrara, G. Pradisi, A. Sagnotti, Y. Stanev and T. Vergata,
hep-th/9607105. 

\bibitem{KLT}
V. Kaplunovsky, J. Louis and S. Theisen, Phys. Lett. {\bf B 357} (1995) 71,
hep-th/9506110.



\bibitem{KLM} A. Klemm, W. Lerche and P. Mayr, \PL{B 357}{95}{313}, 
hep-th/9506112.

\bibitem{AGNT}
I. Antoniadis, E. Gava, K. S. Narain and T. R. Taylor, Nucl. Phys. {\bf B 455}
(1995) 109, hep-th/9507115.

\bibitem{C}
G. Curio, Phys. Lett. {\bf B 366} (1996) 131, hep-th/9509042;
Phys. Lett. {\bf B 368} (1996) 78, hep-th/9509146.


\bibitem{CCLMR}
G. L. Cardoso, G. Curio, D. L\"ust, T. Mohaupt and S.--J. Rey, Nucl. Phys.
{\bf B 464} (1996) 18, hep-th/9512129.


\bibitem{CCLM}
G. L. Cardoso, G. Curio, D. L\"ust and T. Mohaupt, hep-th/9603108,
to appear in Phys. Lett {\bf B}.

\bibitem{WKLL} B. de Wit, V. Kaplunovsky, J. Louis and D. L\"ust,
\NP{B 451}{95}{53}, hep-th/9504006.

\bibitem{AFGNT}
I. Antoniadis, S. Ferrara, E. Gava, K.S. Narain and
T.R. Taylor, \NP{B 447}{95}{35}, hep-th/9504034.


\bibitem{HM}
J. A. Harvey and G. Moore, Nucl. Phys. {\bf B 463} (1996) 315, hep-th/9510182.

\bibitem{FLST} S. Ferrara, D. L\"ust, A. Shapere and S. Theisen, 
\PL{B 225}{89}{363}.



\bibitem{CLM1} G. L. Cardoso, D. L\"ust and T. Mohaupt, \NP {B 432}{94}{68},
hep-th/9405002.

\bibitem{COS}
M. Cvetic, B. Ovrut and  W. Sabra,
Phys. Lett. {\bf B 351} (1995) 173, hep-th/9502144.



\bibitem{MS}
P. Mayr and S. Stieberger, Phys. Lett. {\bf B 355} (1995) 107, hep-th/9504129.



\bibitem{FP} S. Ferrara and A. Van Proeyen, Class. Quantum Grav. {\bf 6} (1989)
L243.

\bibitem{CDFP} A. Ceresole, R. D'Auria, S. Ferrara and A. Van Proeyen,
\NP{B 444}{95}{92}, hep-th/9412200.




\bibitem{CLM2} G. L. Cardoso, D. L\"ust and T. Mohaupt, \NP {B 450}{95}{115},
hep-th/9412209.


\bibitem{BKKM}
P. Berglund, S. Katz, A. Klemm and  P. Mayr, hep-th/9605154.


\bibitem{K2}
T. Kawai, hep-th/9607078.

\bibitem{DVV}
R. Dijkgraaf, E. Verlinde and H. Verlinde, hep-th/9607026;\\
R. Dijkgraaf, G. Moore, E. Verlinde and H. Verlinde, hep-th/9608096.


\bibitem{K1}
T. Kawai, Phys. Lett. {\bf B 371} (1996) 59, hep-th/9512046.


\bibitem{CF}
P. Candelas and A. Font, hep-th/9603170.

\bibitem{LSTY}
J. Louis, J. Sonnenschein, S. Theisen and S. Yankielowicz, hep-th/9606049.


\bibitem{AFIQ}
G. Aldazabal, A. Font, L.E. Ib\'a\~nez and F. Quevedo, Nucl. Phys. {\bf B 461}
(1996) 85, hep-th/969510093.




\bibitem{GSW} M. Green, J. Schwarz and P. West, \NP{B 254}{85}{327}.


\bibitem{GN2}
V. A. Gritsenko and  V. V. Nikulin, alg-geom/9603010.

\bibitem{A}
P. Aspinwall, Phys. Lett. {\bf B 371} (1996) 231, hep-th/9511171.

\bibitem{AL}
P. Aspinwall and J. Louis, Phys. Lett. {\bf B 369} (1996) 233, hep-th/9510234.



\bibitem{GN1}
V. A. Gritsenko and V. V. Nikulin, alg-geom/9510008.


\bibitem{DKL} L. Dixon, V. S. Kaplunovsky and J. Louis, \NP {B 329}{90}{27}.

\bibitem{AGN} I. Antoniadis, E. Gava,  and K. S. Narain,
\PL {B 283}{92}{209},
hep-th/9203071; \NP {B 383}{92}{109}, hep-th/9204030.


\bibitem{EOTY}
T. Eguchi, H. Ooguri, 
A. Taormina and S.--K. Yang, Nucl. Phys. {\bf B 315} (1989) 193.


\bibitem{KYY}
T. Kawai, Y. Yamada
and S.-K. Yang, Nucl. Phys. {\bf B 414} (1994) 191, hep-th/9306096.


\bibitem{Kir} E. Kiritsis, C. Kounnas, P. Petropoulos and J. Rizos,
hep-th/9608034; hep-th/9606087; hep-th/9605011.


\bibitem {HKTY}
S. Hosono, A. Klemm, S. Theisen and S.-T. Yau, Commun. Math. Phys. {\bf 167}
(1995) 301.


\bibitem{FKLZ} S. Ferrara, C. Kounnas, D. L\"ust and F. Zwirner,
Nucl. Phys. {\bf B 365} (1991) 431.

\bibitem{VAFA}  C. Vafa, Nucl. Phys. {\bf B 447} (1995) 261, hep-th/9505023.


\bibitem{NEU} C. D. D. Neumann, hep-th/9607029.




\bibitem{I}
J. Igusa, Amer. J. Math {\bf 84} (1962) 175, {\bf 86} (1964) 392,
{\bf 88} (1966) 817.


\bibitem{EZ}
M. Eichler and D. Zagier, 
{\it The Theory of Jacobi Forms}, Birkh\"auser (1985).





\bibitem{LUTHE} D. L\"ust and S. Theisen, ``Lectures on String Theory",
Springer Verlag, 1989.

\bibitem{LSW} W. Lerche, A.N. Schellekens and N. Warner,
Phys. Rep. {\bf 177} (1989) 1.




\end{thebibliography}
\end{document}